\documentclass[preprint,12pt]{elsarticle}

\usepackage{amssymb}
\usepackage{amsmath}

\usepackage{color}

\journal{European Journal of Mechanics - B/Fluids}

\begin{document}

\begin{frontmatter}



\title{Curvature-based energy spectra revealing flow regime changes in Rayleigh-Bénard convection} 


\author[]{Michael~Mommert\corref{cor1}\fnref{label1}} 
\author[]{Philipp~Bahavar\fnref{label1}} 
\author[]{Robin~Barta\fnref{label1,label2}} 
\author[]{Christian~Bauer\fnref{label1}} 
\author[]{Marie-Christine~Volk\fnref{label1,label2}} 
\author[]{Claus~Wagner\fnref{label1,label2}} 

\cortext[cor1]{michael.mommert@dlr.de}

\affiliation[label1]{organization={Department Ground Vehicles, Institute of Aerodynamics and Flow Technology, German Aerospace Center},
            addressline={Bunsenstr. 10}, 
            postcode={37073}, 
            city={Göttingen},
            country={Germany}}
\affiliation[label2]{organization={Institute of Thermodynamics and Fluid Mechanics, Technische Universität Ilmenau},
			postcode={98684},
			city={Ilmenau},
			country={Germany}}

\begin{abstract}
We use the local curvature derived from velocity vector fields or particle tracks as a surrogate for structure size to compute curvature-based energy spectra.
An application to homogeneous isotropic turbulence shows that these spectra replicate certain features of classical energy spectra such as the slope of the inertial range extending towards the equivalent curvature of the Taylor microscale.
As this curvature-based analysis framework is sampling based, it also allows further statistical analyses of the time evolution of the kinetic energies and curvatures considered.
The main findings of these analyses are that the slope for the inertial range also appears as a salient point in the probability density distribution of the angle of the vector comprising the two time evolution components.
This density distribution further exhibits changing features of its shape depending on the Rayleigh number.
This Rayleigh number evolution allows to observe a change in the flow regime between the Rayleigh numbers $10^6$ and $10^7$.
Insight into this regime change is gathered by conditionally sampling the salient time evolution behaviours and projecting them back into physical space. 
Concretely, the regime change is manifested by a change in the spatial distribution for the different time evolution behaviours.
Finally, we show that this analysis can be applied to measured Lagrangian particle tracks.
\end{abstract}


\end{frontmatter}


\section{Introduction}
\label{sec:intro}

The energy spectrum is one of the classic tools for analysing turbulent flows, as it reveals the energy cascade described by \citet{Richardson1922} and allows to observe the extent of scaling laws, such as the one for the inertial range introduced by \citet{Kolmogorov1991}.
Typically, a Fourier transform is used to compute the relevant relationship of kinetic energy to certain wave numbers.
However, the inherently periodic nature of the Fourier transform leads to artefacts when applied to flows in non-periodic domains such as the cubic Rayleigh-Bénard convection cell studied in this paper. 
For this reason, there are a number of mitigation approaches for spectral solvers that rely on the Fourier transform.
Examples include windowing methods \citep{Schlatter2005} or Fourier continuation techniques \citep{Fontana2020}.
In contrast to the use of extensions to the Fourier transform, we pursue an approach for our analysis which completely circumvents these problems as it is based on local samples.
In more detail, we sample the kinetic energy and (path line) curvature throughout the investigated domain, which enables us to calculate a spectrum of mean kinetic energies for each curvature.
Unlike the wavenumber, the curvature is a local property, which allows the statistical analysis without the implicit requirement of a complete, periodic flow domain.
Regarding the calculation of the curvature of the velocity vector field, \citet{Theisel1996} gives a definition of the curvature of stationary vector fields. A definition regarding the unsteady nature of flows can be found in the work of \citet{Braun2006} for Lagrangian particle tracks, which is transferable to an Eulerian definition of the flow domain. \citet{Braun2006} also describe a trend towards higher curvatures with higher Reynolds numbers, a feature expected for a quantity acting as a measure of structure size.
This curvature measure has been subject of a number of studies, see e.g. \citep{Alards2017,Perven2021,Hengster2024}.
While \citet{Perven2021} investigate the distribution of mean curvatures with a boundary layer, \citet{Alards2017} show that the scaling coefficients of the probability density function of curvature in Rayleigh-Bénard convection (RBC) are the same as for homogeneous isotropic turbulence (HIT) as long as the boundary layers are excluded.
\citet{Hengster2024} confirm this and reveal an anisotropic behaviour of the Cartesian curvature components associated with the anisotropic nature of the RBC flow. 

Here, we use the curvature in conjunction with the kinetic energy to study RBC flows. Specifically, we investigate how a change in flow regime manifests itself for the proposed analysis methods in order to shed more light on the characteristics of the regimes.
RBC flows can be classified into flow regimes depending on their main control parameters, the Rayleigh number $\mathrm{Ra}$ and the Prandtl number $\mathrm{Pr}$.
Because of the importance for flows occurring at large scales in nature, a recent focus of investigation has been on the transition from the so-called classical regime to the ultimate regime, which is characterised by strong forcing ($\mathrm{Ra}\gtrsim 10^{11}$) and enhanced heat transfer due to fully turbulent boundary layers. A summary of these efforts is given by \citet{Lohse2024a}.
However, since these Rayleigh numbers are difficult to access both numerically and experimentally, we consider a regime transition within the classical regime for the present study.
Specifically, we consider a Rayleigh number range $5\times10^5 \leq \mathrm{Ra} \leq 10^9$ at a Prandtl number of $0.7$.
This range was chosen because it covers the transition from boundary layer dominated to bulk dominated thermal dissipation according to the Grossmann-Lohse theory \citep{Grossmann2000, Stevens2013} within the classical regime.
It also coincides with the expected transition between soft and hard turbulence observed by \citet{Heslot1987} and \citet{Castaing1989}.
Although this accessible range of control parameters has already been extensively studied, the recent study by \citet{Castaing2024} shows the potential to refine existing theories of regime transitions.

The methodology for investigating the described parameter range is introduced in section~\ref{sec:curv-spec-method}. First, we use a generic flow pattern (section~\ref{sec:curv-spec-conversion}) and a HIT dataset provided by the Johns Hopkins Turbulence Databases (JHTDB) (section~\ref{sec:curv-spec-hit}) to explore its capabilities.
The method is then applied to RBC cases with different $\mathrm{Ra}$ in section~\ref{sec:rbc}. 
In addition to discussing the Rayleigh number range $5\times10^5 \leq \mathrm{Ra} \leq 10^9$, we also apply the analysis framework to Lagrangian particle tracking (LPT) data for $\mathrm{Ra} = 2.5 \times 10^9$ and $\mathrm{Pr} = 7$ to investigate its suitability for experimental data in section~\ref{sec:ptv}.

\section{Methodology}\label{sec:curv-spec}

This section aims to introduce the basics of curvature-based spectra and to explore their characteristics for a segment of the JHTDB HIT dataset ($\mathrm{Re}_\lambda = 433$), on which the more extensive analyses are introduced.

\subsection{Curvature-based energy spectra}\label{sec:curv-spec-method}
For the proposed sampling-based investigation of the relationship between kinetic energy and curvature, both quantities must be calculated for each sampling point within the domain. In the case of the dimensionless kinetic energy $E_\mathrm{kin}$, this is a simple dependence on the velocity vector $\boldsymbol{u}$:
\begin{align}
 E_\mathrm{kin}=\frac{1}{2}\boldsymbol{u}\cdot\boldsymbol{u}.
\end{align}

As mentioned in the introduction, the curvature $\kappa$ \citep{Braun2006} is used as a representation of the structure size:
\begin{align}
 \kappa = \| \boldsymbol{\kappa}\| = \left\| \frac{\boldsymbol{u} \times ( \frac{\partial \boldsymbol{u}}{\partial t} + (\boldsymbol{u} \cdot \nabla )\boldsymbol{u}) }{\|\boldsymbol{u}\|^3} \right\|.
\end{align}

In this definition, the curvature is the magnitude of a curvature vector $\boldsymbol{\kappa}$ based on the cross product of the velocity of a fluid parcel and its acceleration, i.e. the material time derivative of the velocity. The curvature vector is therefore perpendicular to the plane in which the fluid flow is curved and its magnitude is equivalent to the reciprocal value to the local radius of curvature.
For the purposes of this study, the first step in analysing the relationship between $E_\mathrm{kin}$ and $\kappa$ is to determine their joint distribution based on the relative incidences. For this, the samples are collected in logarithmically spaced bins of curvature $B_i^\kappa = [\kappa_{i-1}, \kappa_i[$ and kinetic energy $B_j^E = [E_{\mathrm{kin}\,j-1}, E_{\mathrm{kin}\,j}[$. The respective relative incidence $f_{ij}$ is defined as

\begin{align}\label{eq:rel_inc}
 f_{ij} = \frac{ \sum_{s=1}^N \mathbb{I}(\kappa_s \in B_i^\kappa, \; E_{\mathrm{kin}\,s} \in B_j^E)\, V_s}{N \sum_{s=1}^N V_s},
\end{align}

where $s$ is the running index of the samples, $N$ is the total number of samples, $V_s$ is the volume each sample represents in the simulation domain used as weight, and $\mathbb{I}$ is the binary indicator function that returns 1 for samples that belong to the respective bin.

To derive a spectral function $E_\mathrm{kin}(\kappa)$ from this joint distribution, weighted averages of $E_\mathrm{kin}$ are calculated for each $\kappa$ bin:
\begin{align}\label{eq:spec_line}
 E_\mathrm{kin}(\kappa) = \langle E_\mathrm{kin} \rangle_{B_i^\kappa} = \frac{ \sum_{s=1}^N E_{\mathrm{kin}\,s}\, \mathbb{I}(\kappa_s \in B_i^\kappa)\, V_s}{\sum_{s=1}^N \mathbb{I}(\kappa_s \in B_i^\kappa)\, V_s}.
\end{align}

\subsection{Wavenumber-curvature relation}\label{sec:curv-spec-conversion}

\begin{figure}[h]
\centering
\includegraphics[trim={0 15 0 0},clip]{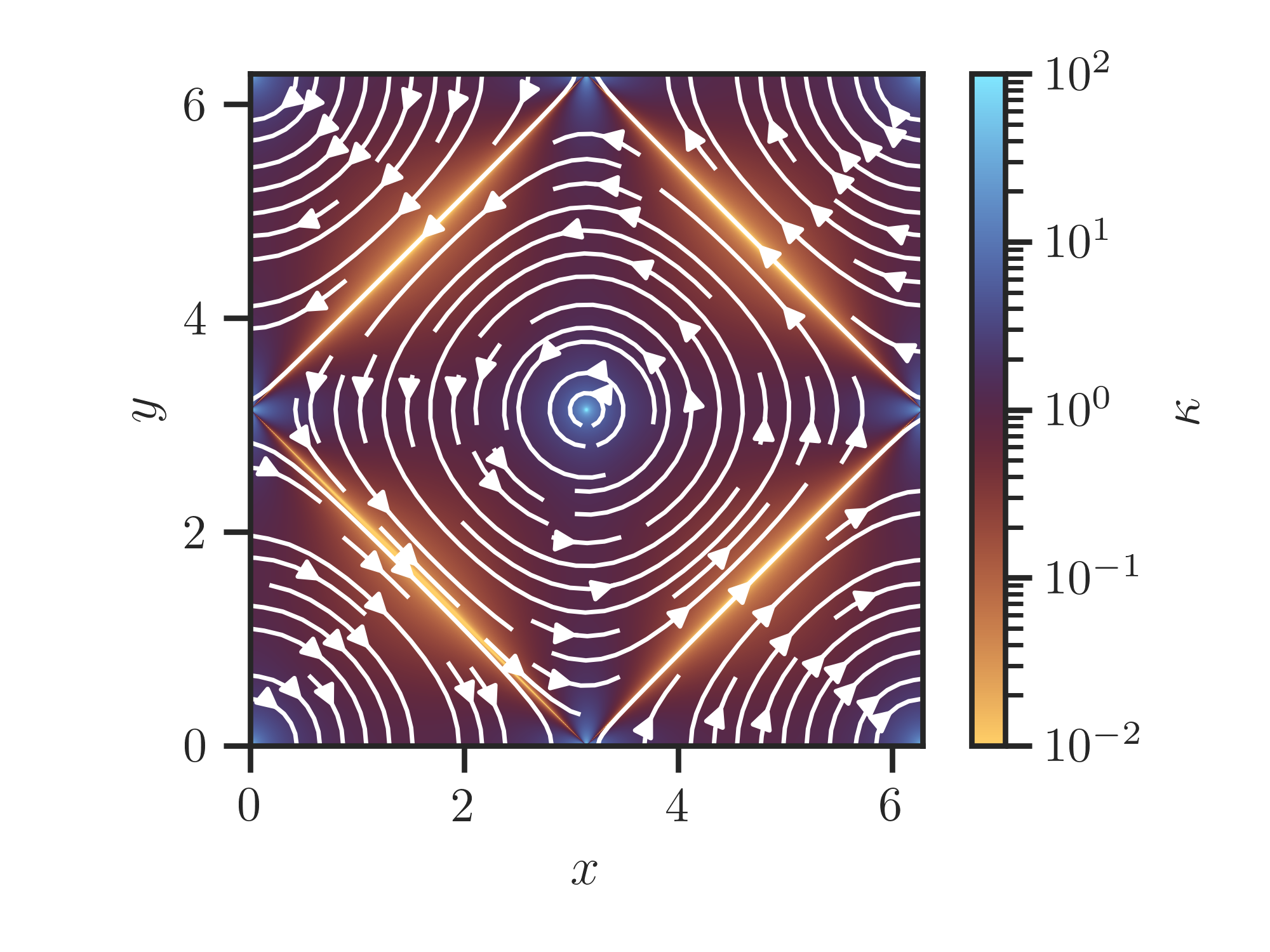}
\caption{Generic planar flow representing a wavenumber of $k=1$. The respective curvatures $\kappa$ are coded by colour.}\label{fig:gen_vort}
\end{figure}

The use of curvature as a measure of flow structure sizes naturally raises the question of how curvature is related to the traditionally used wavenumber of an FFT. 
For this reason we consider a generic, stationary velocity field

\begin{align}
 u_x(x,y) &= \sin(y) \\
 u_y(x,y) &= -\sin(x),
\end{align}

that represents a wavenumber of $k=1$ in a planar\footnote{We consider a planar domain as sufficient for this investigation since each curvature vector describes a plane in which the rotation is considered.}, periodic domain with $x$ and $y$ ranging from $0$ to $2\pi$.
A streamline visualisation of the resulting vector field with the respective curvatures coded by colour is displayed in Figure~\ref{fig:gen_vort}. It reveals the characteristic that a single wavenumber is associated with a range of curvatures, including very large values in the centre of the circulation and at the stagnation points, as well as very small values at the interfaces between the synthetic circulations.

\begin{figure}[h]
\centering
\includegraphics{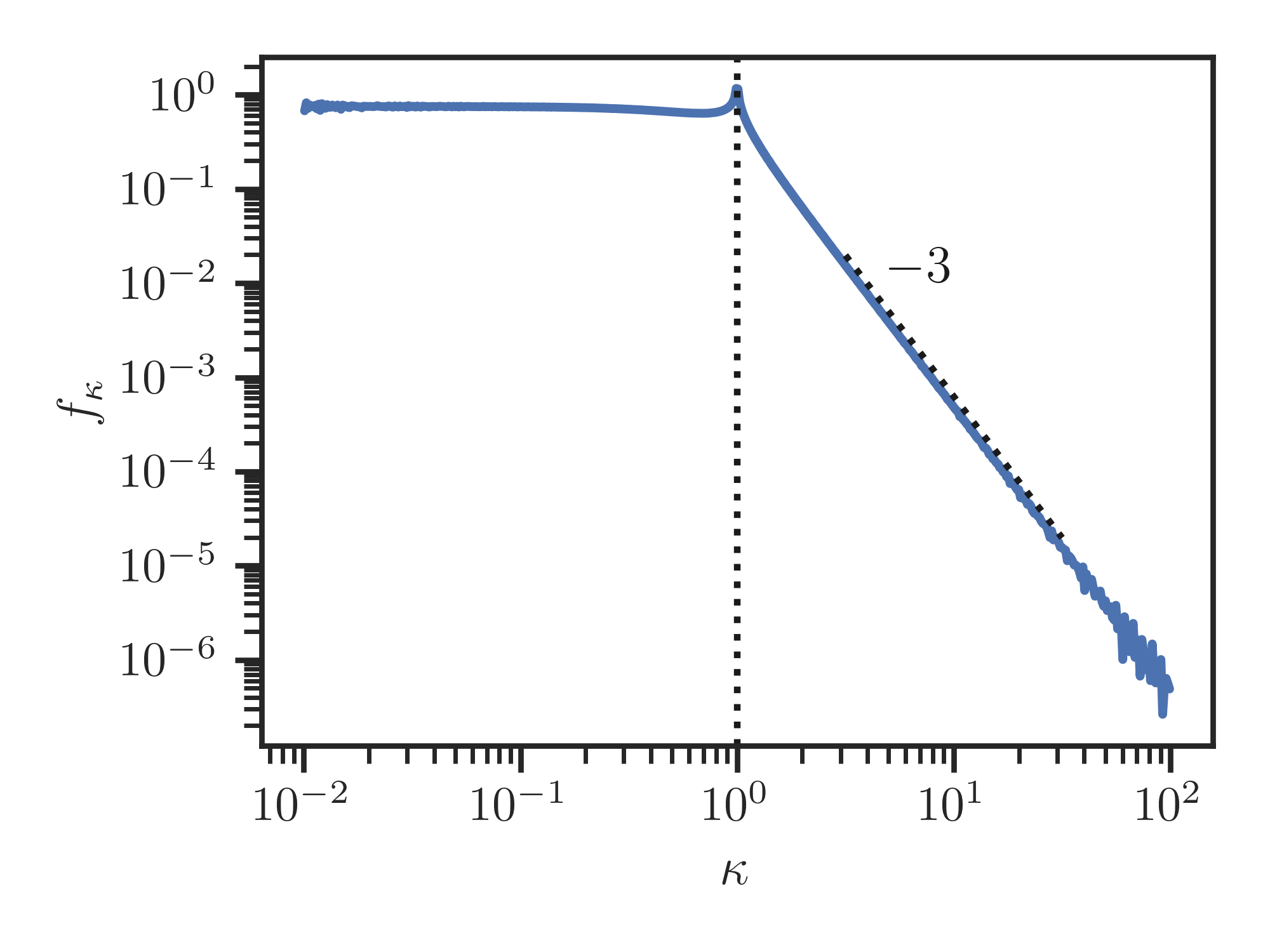}
\caption{Distribution of incidence of the curvatures for a synthetic planar velocity field with wavenumber k=1 displayed in Figure~\ref{fig:gen_vort}.}\label{fig:gen_vort_distr}
\end{figure}

Figure~\ref{fig:gen_vort_distr} shows the corresponding distribution of incidences of $\kappa$ for spatially homogeneous sampling of this generic velocity vector field. It reveals relatively high constant values for small curvatures up to a curvature of exactly $\kappa=1$, which is the distinct modal value of this distribution. The range from the modal value towards large curvatures shows an exponential decay with a slope of $-3$ which results from the geometry of the circular flow present in the centre of the circulation and near the stagnation points. The derivation of this slope is given in \ref{app:curv_distr_circ}. Overall, these results indicate that a wavenumber of $k=1$ is associated with a curvature of $\kappa=1$ for a domain size of $2\pi$. However, this relationship does not imply equality, which means that the distribution of a single wavenumber over a range of curvatures must be taken into account for the subsequent analyses.  

\subsection{Properties of curvature-based spectra for JHTDB HIT}\label{sec:curv-spec-hit}

Having established the relationship between wavenumber and curvature, the next step is to explore the properties of the method for a periodic reference case. To do this, we applied the method described above to a case of forced homogeneous isotropic turbulence \citep{Minping2012} provided by the JHTDB \citep{Perlman2007, Li2008}.

\begin{figure}[h]
    \begin{center}
        \begin{tabular}{cc}
            {a)} & {b)} \\
            \includegraphics[width=0.45\textwidth]{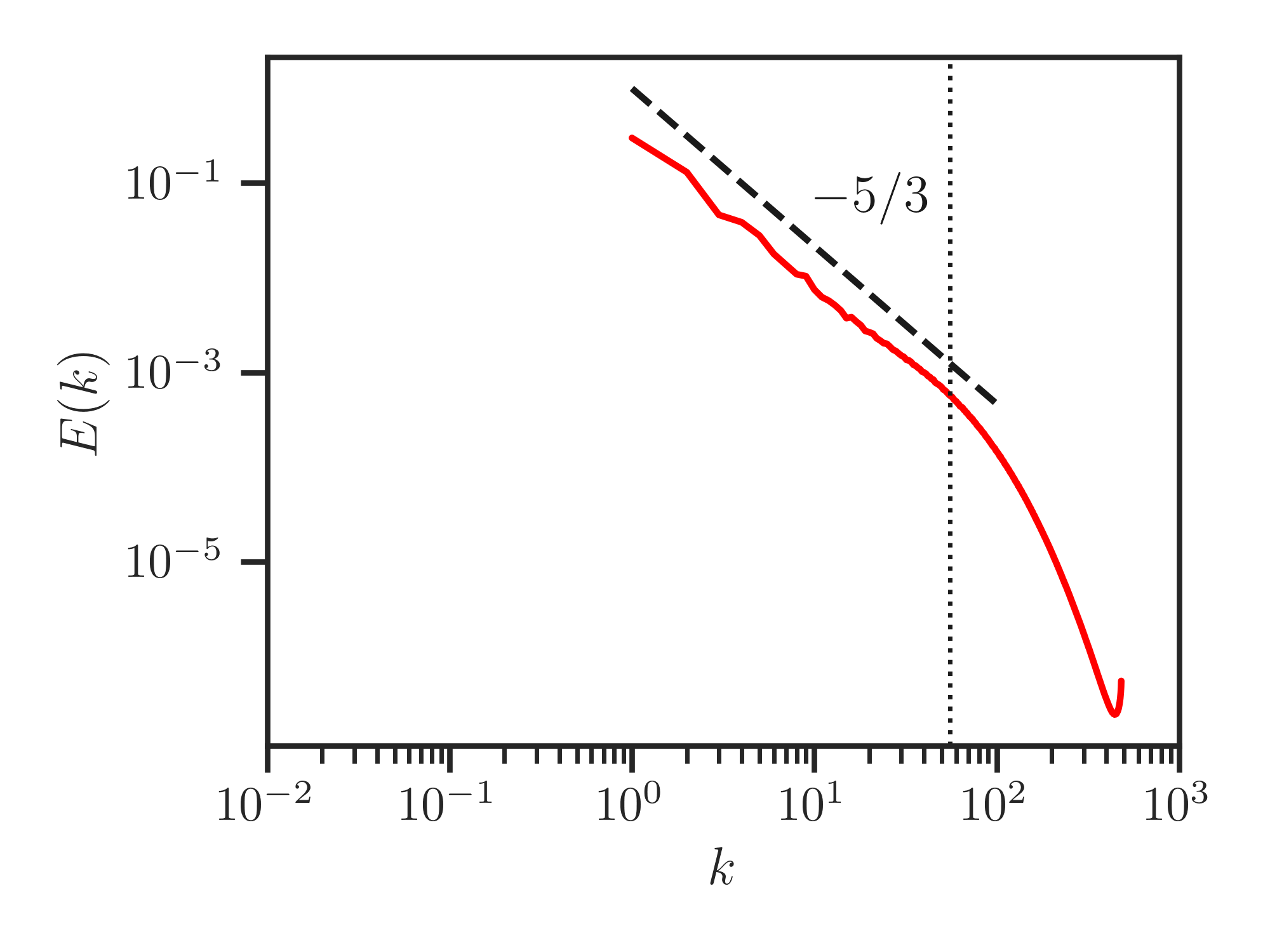} &
            \includegraphics[width=0.55\textwidth]{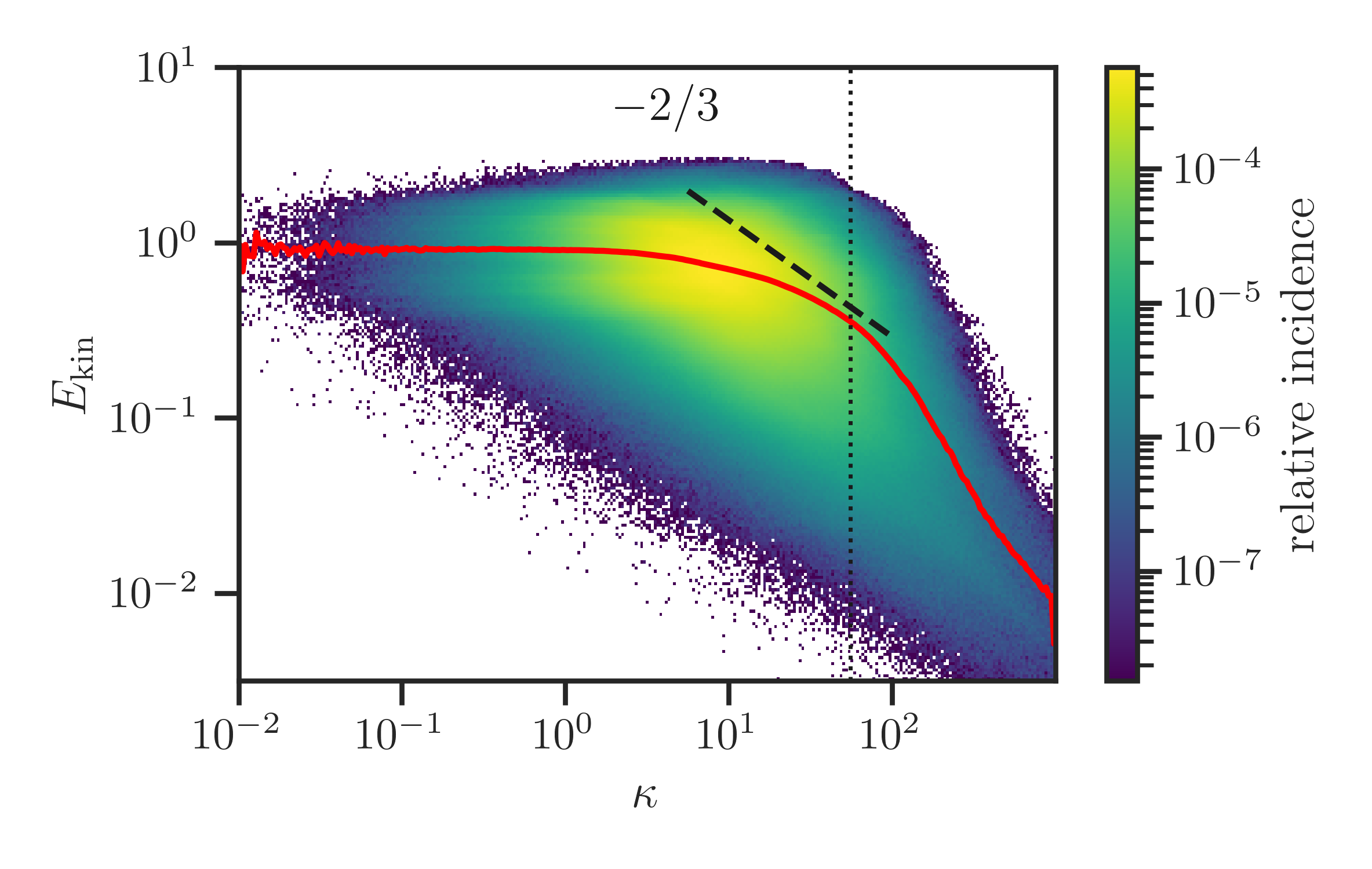}
        \end{tabular}
        \caption{a) Classical energy spectrum for the JHTDB HIT flow provided by \citep{Li2008}. b) Curvature-based energy spectrum (red line) based on Equation~\ref{eq:spec_line} for a segment of the JHTDB HIT dataset. Dashed lines indicate the respective slopes for the inertial range. The dotted lines mark the abscissa values associated with the Taylor microscale $\lambda$. In the background, the relative incidences $f_{ij}$ of the joint distribution from which it originates is displayed.}\label{fig:spec_HIT}
    \end{center}
\end{figure}

The information provided for this flow by the JHTDB includes the classical energy spectrum shown in Figure~\ref{fig:spec_HIT}a), as well as the Taylor length scale of $\lambda = 0.118$ (represented by the dotted lines in Figure~\ref{fig:spec_HIT}). In comparison, the curvature-based energy spectrum shown in Figure~\ref{fig:spec_HIT}b) was calculated using the method presented in section~\ref{sec:curv-spec-method} for a $32\times32\times32$ grid segment over $2000$ time steps separated by $\delta t = 0.002$. All required gradients are calculated by second order accurate finite differences with a central scheme for interior points and a one-sided scheme at the boundaries \citep{Fornberg1988}.

When comparing the two spectra, it is important to note that the inherent dimensionality of the kinetic energy $E_\mathrm{kin}(\kappa)$ does not include a length dimension stemming from integration, as is the case for $E(k)$.
Therefore, the expected slope for the inertial range of the curvature spectrum can be derived by its sole dependence on $k$ or in this case $\kappa$ and the turbulent dissipation $\epsilon$ 

\begin{align}
  E_\mathrm{kin} &\propto \epsilon^i \kappa^j, \\
  \frac{\mathrm{length}^2}{\mathrm{time}^2} &\propto
  \left(\frac{\mathrm{length}^2}{\mathrm{time}^3}\right)^i \left(\frac{1}{\mathrm{length}}\right)^j, \\
  \Rightarrow i &= 2/3\,;\ j= -2/3,
\end{align}

which gives an expected slope of $-2/3$ for the inertial range of the energy cascade in a curvature-based energy spectrum.

Comparing the two types of energy spectra in Figure~\ref{fig:spec_HIT} shows that the curvature spectrum (b) extends over a wider range of the respective abscissa, since the theoretical limits of the classical spectrum (a), namely the domain size and the Nyquist wavenumber, do not apply to the curvature.
Accordingly, the curvature spectrum exhibits an asymptotic behaviour for $E_\mathrm{kin} \approx 1$ for small curvatures.
Together with the relationship between wavenumber and curvature displayed in Figure~\ref{fig:gen_vort_distr}, this results in a rather narrow $\kappa$ range with the slope of $-2/3$.
However, the behaviour in conjunction with the Taylor microscale $\lambda$ appears equal for both types of spectra. In fact, the wavenumbers or curvatures associated with $\lambda$ (marked by dotted lines) indicate the structure size limit on which viscosity begins to play a significant role. Therefore, both types of spectra show steeper decays than for the inertial range beyond this limit. 
Another difference emerges for very large wavenumbers or curvatures. While these are completely suppressed by viscosity for the classical case, the relationship presented in Figure~\ref{fig:gen_vort_distr} shows, that energy for a certain wavenumber can be distributed to arbitrarily small curvatures. Consequently, the curvature spectrum extends towards very large curvatures.

This comparison showed that the curvature-based spectrum replicates the characteristic effect of viscosity, namely a steeper decline associated with the Taylor microscale, of the classical spectrum.
While there are also inherent differences, a sampling-based approach also has the advantage that the link to the original data persists and can be used for further analysis.
An example of this is the consideration of the time evolution of individual fluid parcels within the $E_\mathrm{kin}$-$\kappa$ plane, in a visualisation approach similar to the one of \citet{Jimenez2024} for different projection variables.
To find possible prevalent paths of this time evolution, we again rely on binning statistics. For this, we calculate the time evolution vector

\begin{align}\label{eq:time_vec}
 \boldsymbol{U}^* = \begin{bmatrix} \frac {D \kappa / D t}{\log_e(10)\, \kappa} &  \frac {D E_\mathrm{kin} / D t}{\log_e(10)\, E_\mathrm{kin}}    \end{bmatrix}^\top
\end{align}

where $D \phi/D t$ is the material derivative of an arbitrary scalar $\phi$. Note that the normalisation with the respective local values of $E_\mathrm{kin}$ and $\kappa$ as well as the natural logarithm of $10$ results from considering the vector $\boldsymbol{U}^*$ in the double-logarithmic $E_\mathrm{kin}$-$\kappa$ plane\footnote{Quantities $\phi^*$ with an asterisk indicate the representations of $\phi$ within a double-logarithmic plane with base 10.}:

\begin{align}
 \phi^* &= \log_{10}(\phi) ,\\
 \frac{D\phi^*}{Dt} &= \frac{1}{\phi\, \log_e(10)} \frac{D\phi}{Dt}.
\end{align}

Similar to Equations~\ref{eq:rel_inc} and \ref{eq:spec_line}, we calculate the resulting vector field $\boldsymbol{U}^*(\kappa, E_\mathrm{kin})$ by means of bin averages

\begin{align}
 \boldsymbol{U}^*(\kappa, E_\mathrm{kin}) = \langle \boldsymbol{U}^* \rangle_{B_i^\kappa, B_j^E} = \frac{ \sum_{s=1}^N \boldsymbol{U}^*_s\, \mathbb{I}(\kappa_s \in B_i^\kappa, \; E_{\mathrm{kin}\,s} \in B_j^E)\, V_s}{\sum_{s=1}^N \mathbb{I}(\kappa_s \in B_i^\kappa, \; E_{\mathrm{kin}\,s} \in B_j^E)\, V_s}.
\end{align}

\begin{figure}[h]
\centering
\includegraphics{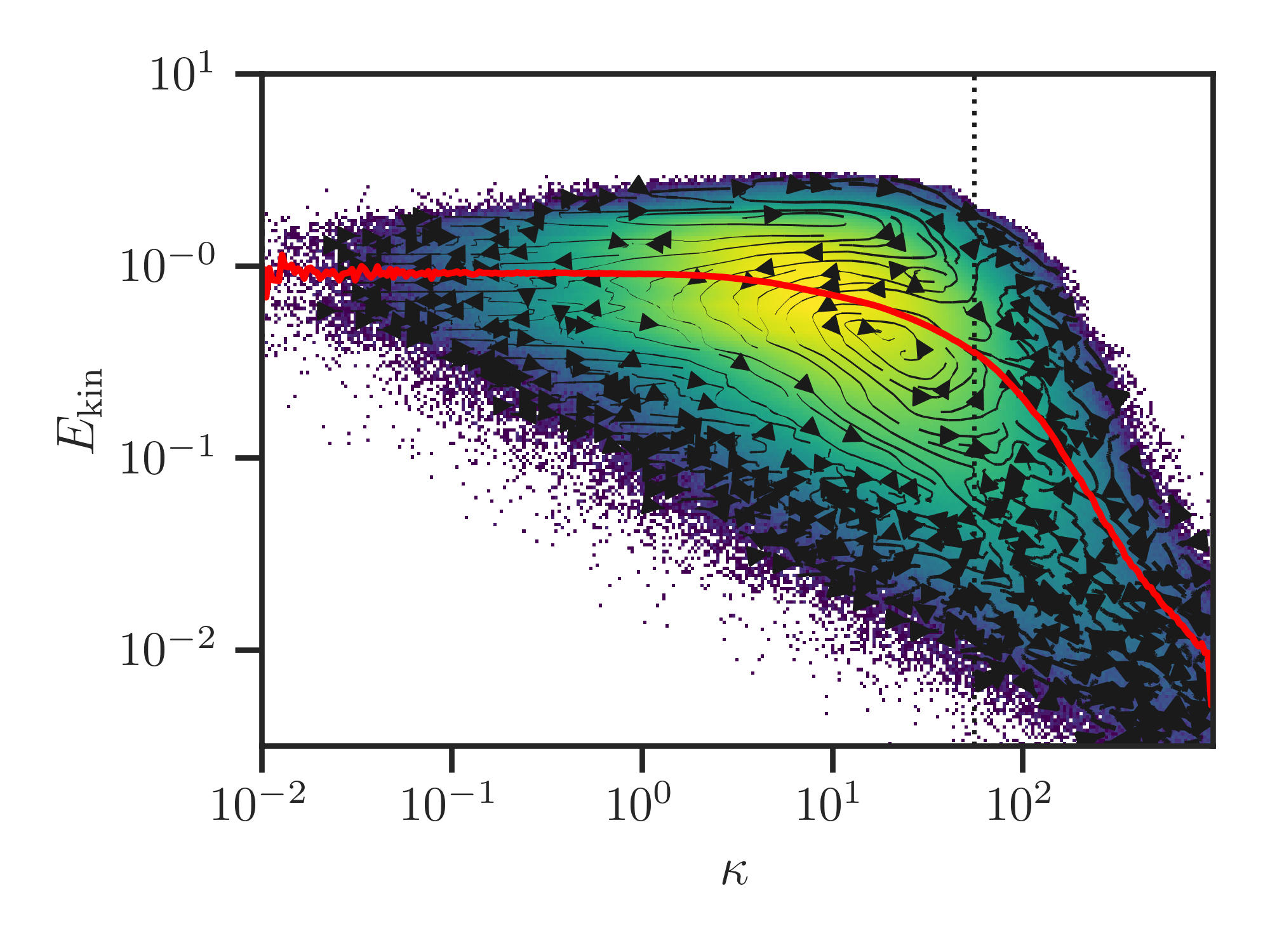}
\caption{Curvature-based energy spectrum for HIT from Figure~\ref{fig:spec_HIT} extended by a streamline visualisation of the mean time evolution vector field $\langle \boldsymbol{U}^* \rangle_{B_i^\kappa, B_j^E}$.}\label{fig:hit_timeevo}
\end{figure}

Figure~\ref{fig:hit_timeevo} shows a streamline visualisation of the resulting vector field of $\boldsymbol{U}^*$ in the $E_\mathrm{kin}$-$\kappa$ plane. Coherent patterns are particularly evident in regions of high relative incidence. However, due to the limited size of the analysed segment of the dataset, this plot is still state-dependent rather than revealing a completely converged mean process.
Nevertheless, the existence of coherent structures of $\boldsymbol{U}^*$ raises the question of whether laws such as the $-2/3$ slope are not only prevalent in a statistical sense, but also for the time evolution.
To investigate this, we first define the angular direction $\theta$ of each vector $\boldsymbol{U}^*$ by

\begin{align}
 \theta = \mathrm{atan2}(U^*_{E_\mathrm{kin}},U^*_{\kappa}) .
\end{align}

In a physical sense, the angle $\theta$ indicates whether a fluid parcel is experiencing a curving ($\theta=0$) or straightening ($\theta=\pm\pi$) of its trajectory, whether it is accelerated ($\theta=\pi/2$) or decelerated ($\theta=-\pi/2$) along its path, or combinations of both.
Further, large magnitudes of $\boldsymbol{U}^*$ represent fast state changes regarding $E_\mathrm{kin}$ and $\kappa$. As the same effective motion within the $E_\mathrm{kin}$-$\kappa$ plane is less likely to be sampled for large $\|\boldsymbol{U}^*\|$, we correct this effect by adding $\|\boldsymbol{U}^*\|$ to the weighting for the densities $p_i$ of each angular bin $B_i^\theta = [\theta_{i-1}, \theta_i[$ with widths $\Delta_i$:

\begin{align}\label{eq:dens_angle}
 p_{i} = \frac{ \sum_{s=1}^N \mathbb{I}(\theta_s \in B_i^\theta)\, V_s\, \|\boldsymbol{U}^*_s\|}{\Delta_i \sum_{s=1}^N V_s \, \|\boldsymbol{U}^*_s\|}.
\end{align}

\begin{figure}[h]
\centering
\includegraphics[width=0.95\textwidth]{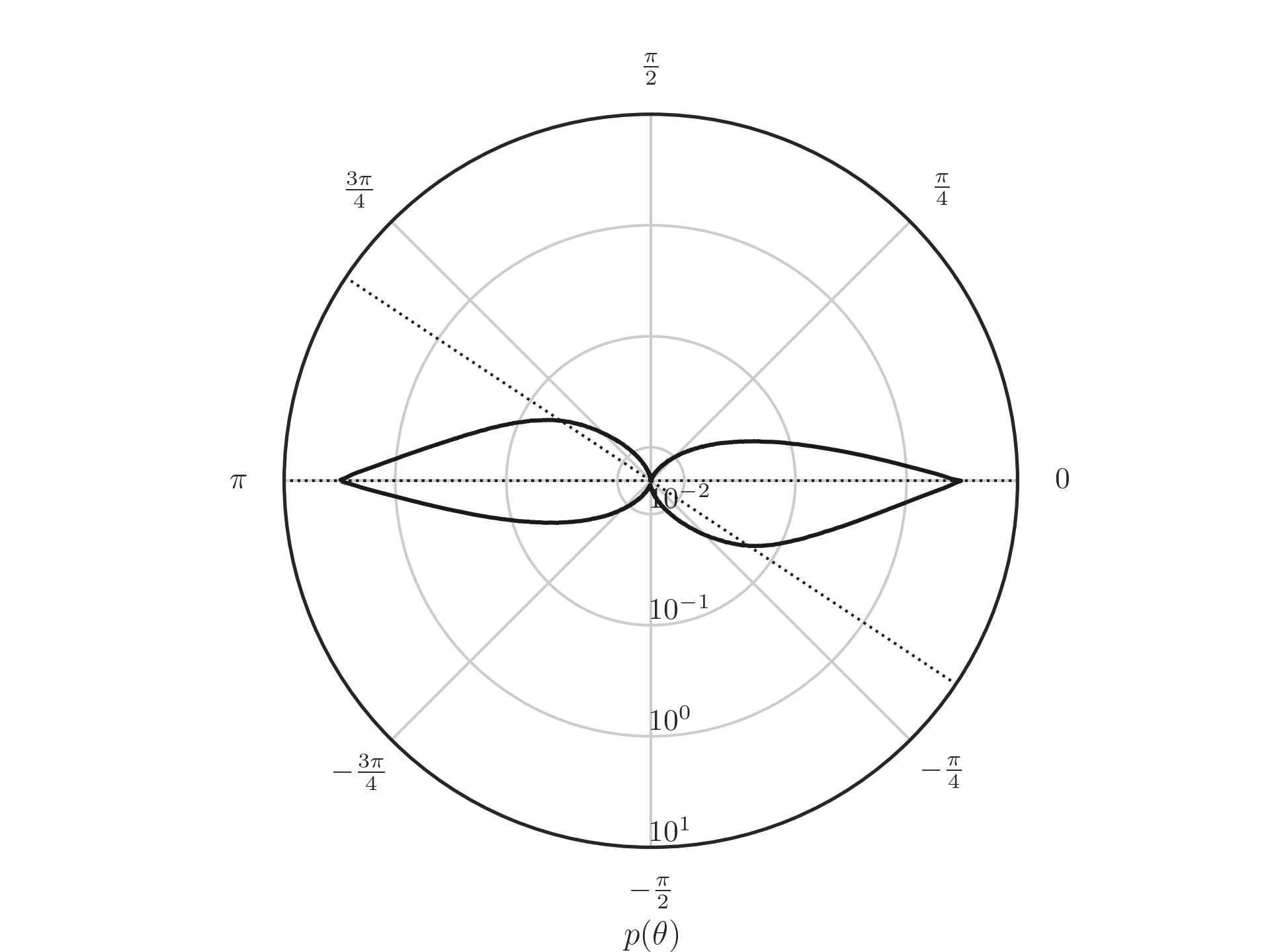}
\caption{Polar density distribution of the angle $\theta$ of the time evolution vector $\boldsymbol{U}^*$ for the JHTDB HIT flow case. The angle associated with the $-2/3$ slope is marked by a dotted line.}\label{fig:hit_slopes}
\end{figure}

The resulting polar density distribution is displayed in Figure~\ref{fig:hit_slopes}. It reveals a bimodal distribution with two peaks at the angles $0$ and $\pm \pi$, corresponding to an evolution of the curvature towards larger or smaller values, respectively, without significant changes in the kinetic energy. Thus, these peaks are associated with ideal inertial behaviour, where the curvature changes of the trajectory of a fluid parcel do not result in any acceleration or deceleration along its path.
Both peaks exhibit shoulders that characterise skews extending to the respective angles associated with the $-2/3$ slope, which is marked by a dotted line. Beyond these angles, the densities decrease more rapidly.
Especially remarkable in this regard is the almost perfect rotational symmetry of the distribution\footnote{see also Figures~\ref{fig:slopes_gath} and \ref{fig:tracks_slopes}}. 
For the flow being in an equilibrium, one would expect both sides, representing increasing (positive $\theta$) and decreasing (negative $\theta$) kinetic energy, to be balanced. 
However, this does not require this rotational symmetry. 
Its existence therefore indicates that the process under which a fluid parcel's trajectory changes its curvature while subject to forcing is symmetric for acceleration and deceleration.

\section{Rayleigh-Bénard data}
\subsection{Rayleigh number variation}\label{sec:rbc}

Using the curvature spectrum analysis introduced for the case of HIT, we now turn to RBC in a cubic domain to study the evolution of its characteristics in the context of flow regime changes.
To do this, we focus on the Rayleigh number dependence for a given Prandtl number of $Pr=0.7$.
The data sets considered for this investigation are listed in Table~\ref{tab:DNS}.
These data sets have been generated by a fourth-order DNS solver that has been validated for a variety of flows. RBC examples include the studies of \citet{Shishkina2007}, \citet{Kaczorowski2010}, \citet{Kaczorowski2013}.
The dimensionless time used for this, refers to units of free-fall time $t_\mathrm{ff}=\sqrt{H/(a g \Delta T)}$, where $H$ is the height of the cubic domain, $a$ the thermal diffusivity, $g$ the gravitational acceleration, and $\Delta T$ the temperature difference between the top and bottom faces of the domain.

As the presented analysis of curvature spectra is based on sampling, we aimed to consider the same number of sampling points, specifically $N\approx 3.2\times10^9$, for each case, except for $\mathrm{Ra}=10^9$ which is also sampled with a smaller frame separation $\delta t$ to ensure correct time derivatives (see \ref{app:stat_sensitivity}). 
This results in a large variation of the considered time intervals between small and large Rayleigh numbers, which is addressed in \ref{app:stat_sensitivity} to show that the discussed statistics are not time interval sensitive and sufficiently converged.

\begin{table}[h]
\centering
\begin{tabular}{c c c c c}
  $\mathrm{Pr}$ & $\mathrm{Ra}$ & $N_x \times N_y \times N_z$ & $N_t$ & $\delta t$ \\
  $0.7$ & $5 \times 10^5$ & $64 \times 64 \times 64$ & $12096$  & $0.01$\\
  $0.7$ & $10^6$ & $96 \times 96 \times 96$ & $3584$  & $0.01$\\
  $0.7$ & $10^7$ & $160 \times 160 \times 160$ & $774$  & $0.01$\\
  $0.7$ & $10^8$ & $384 \times 384 \times 384$ & $56$  & $0.01$\\
  $0.7$ & $10^9$ & $768 \times 768 \times 768$ & $12$ & $0.001$\\
\end{tabular}
\caption{Overview over the investigated RBC cases. The number of available time frames $N_t$ with a separation of $\delta t$ is adjusted, so that approximately $3.2\times 10^9$ overall samples are considered for each case.}\label{tab:DNS}
\end{table}

\begin{figure}[h!]
    \begin{center}
        \begin{tabular}{c c c}
            a) & b) & c)  \\
            \includegraphics[trim={0 25 60 50},clip, height=0.33\textwidth]{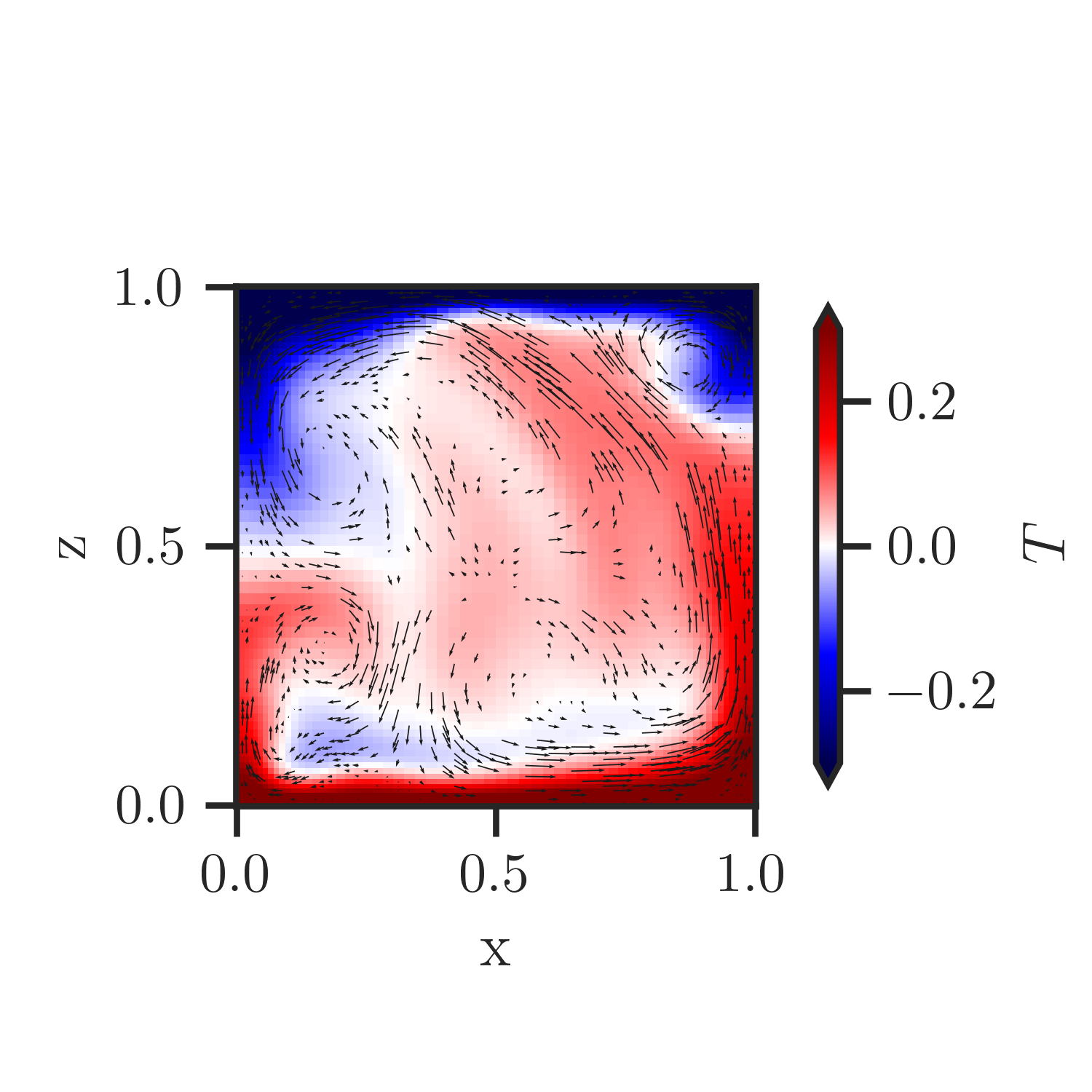} &
            \includegraphics[trim={38 25 60 50},clip, height=0.33\textwidth]{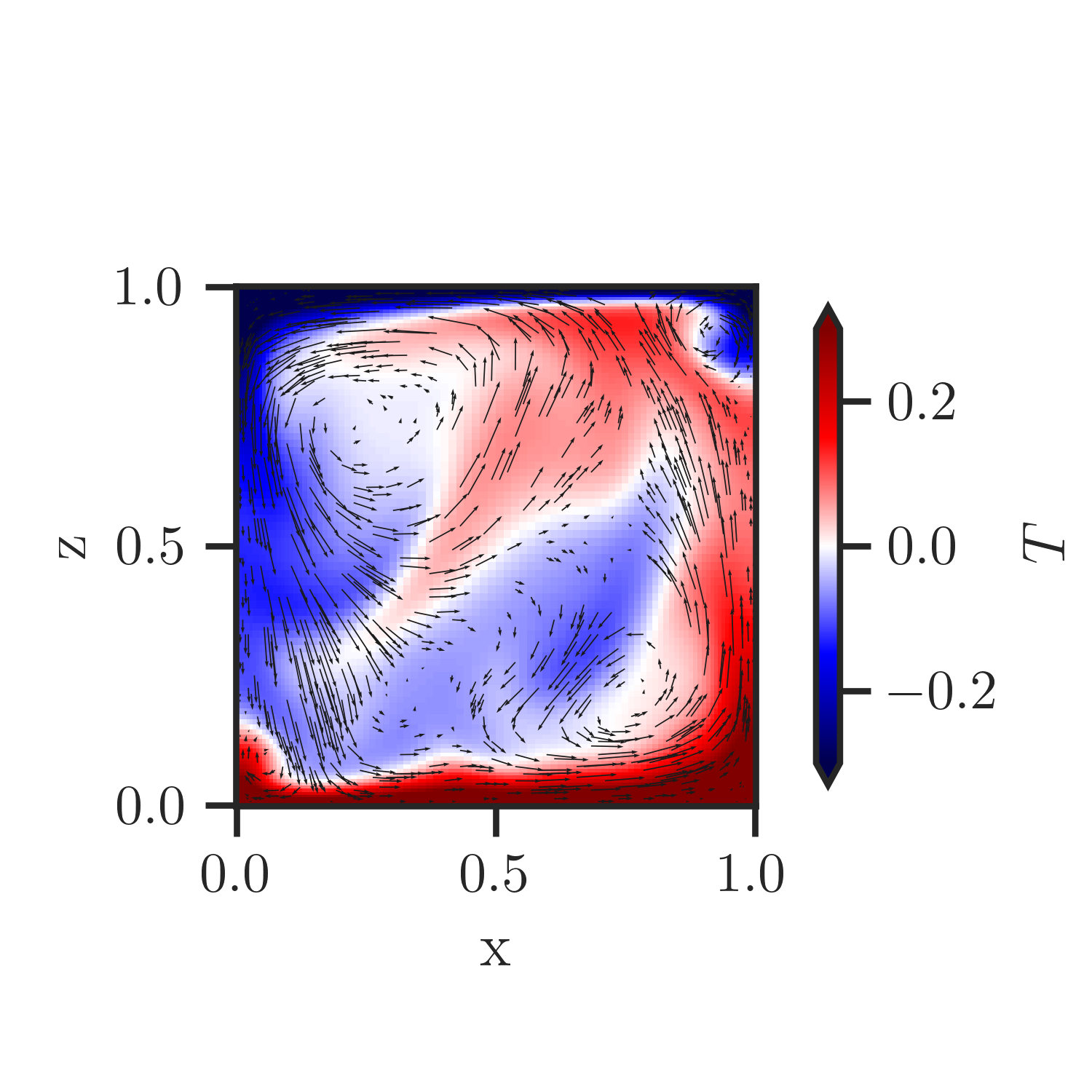} &
            \includegraphics[trim={38 25 60 50},clip, height=0.33\textwidth]{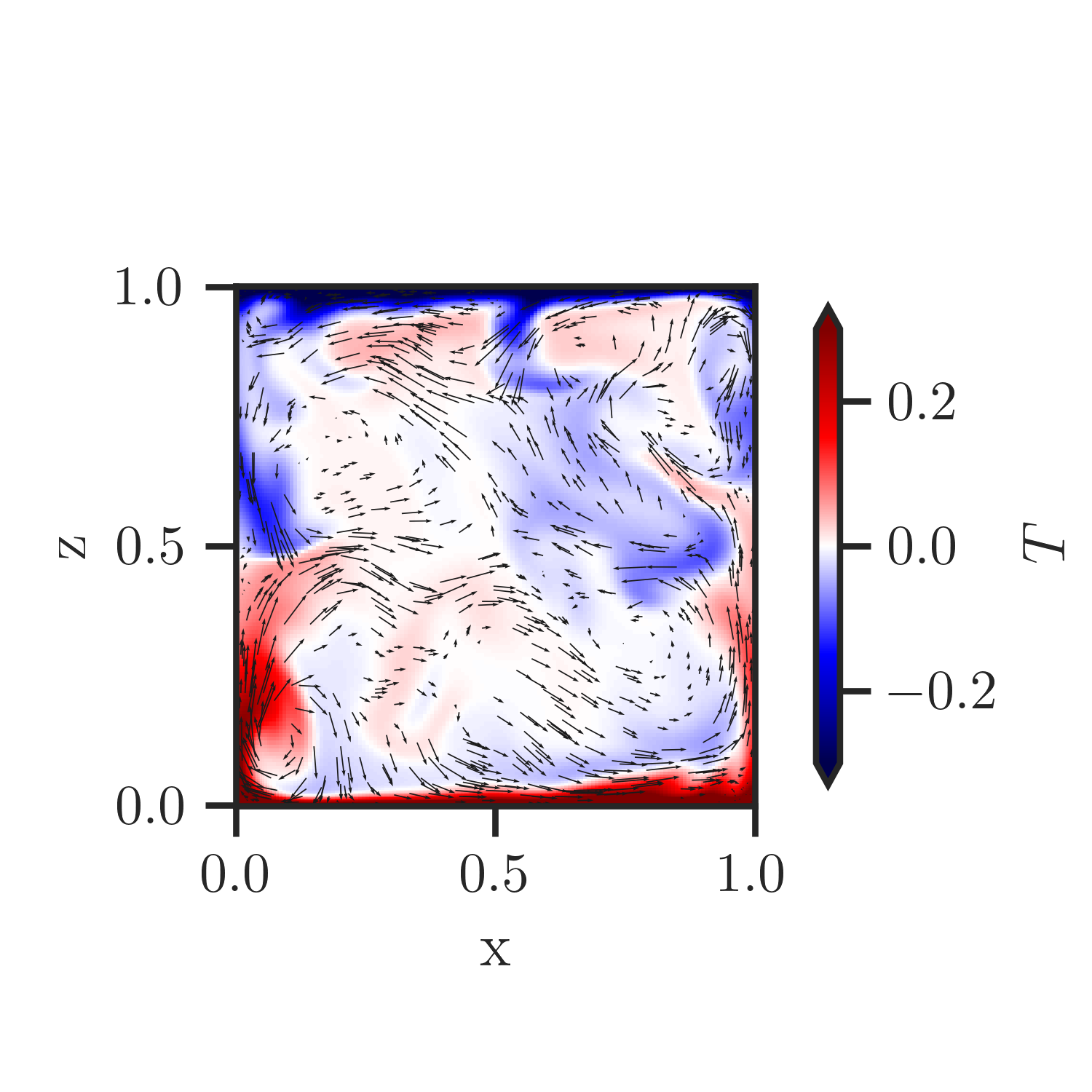}\\
        \end{tabular}
        \begin{tabular}{c c}
            {d)} & {e)} \\
            \includegraphics[trim={0 25 60 50},clip, height=0.33\textwidth]{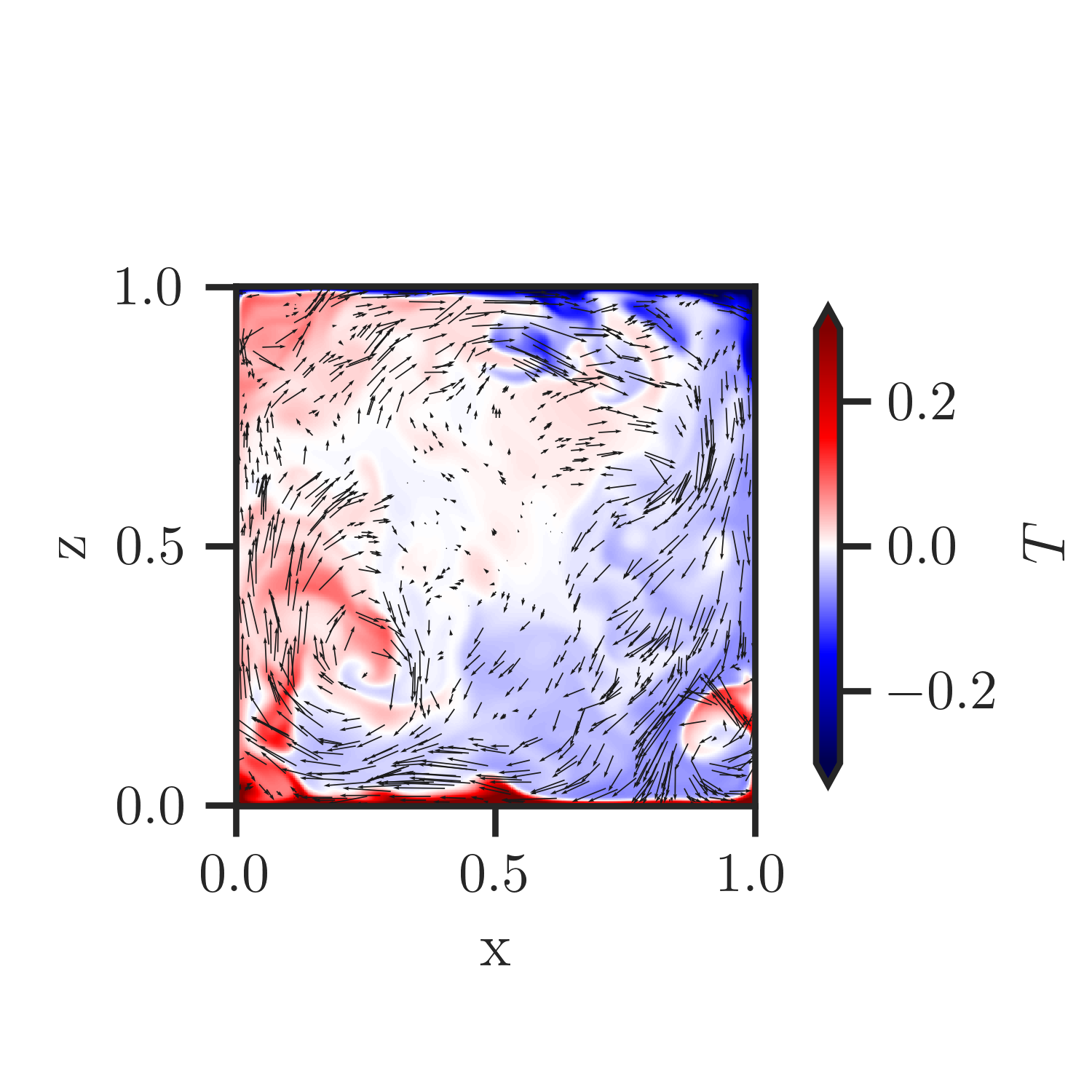} &
            \includegraphics[trim={38 25 5 50},clip, height=0.33\textwidth]{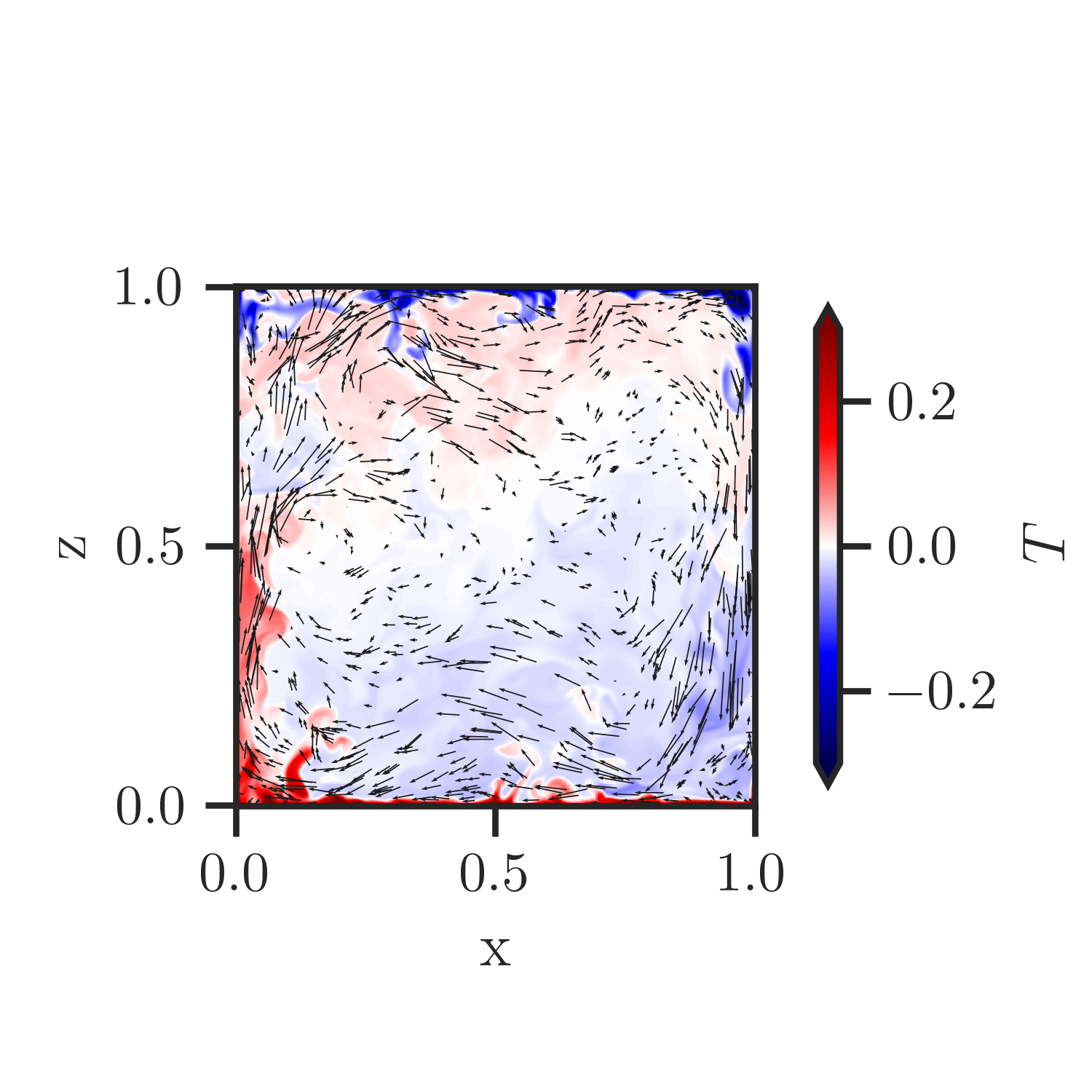}
        \end{tabular}
        \caption{Exemplary fields of the temperature (colour-coded) and the in-plane velocity components (vectors) in a central vertical section of the domain. a) $\mathrm{Ra}=5\times 10^5$ b) $\mathrm{Ra}=10^6$ c) $\mathrm{Ra}=10^7$ d) $\mathrm{Ra}=10^8$ e) $\mathrm{Ra}=10^9$.}\label{fig:exem_fields}
    \end{center}
\end{figure}

As an overview of these cases, Figure~\ref{fig:exem_fields} displays exemplary, instantaneous fields of colour-coded temperatures within a vertical central section of the domain. 
They are complemented by randomly sampled vectors representing the in-plane velocity components. Overall, they show hot and cold structures, namely plumes, detaching from the bottom and top plate, respectively. As it is typical for RBC, these plumes contribute to a large-scale circulation (LSC), which is usually aligned along one of the vertical diagonal planes of the cubic domain. Besides the presence of LSCs, the series of flow visualisations displays the increase in intricacy, i.e. smaller structure sizes, with increasing Rayleigh number.

As a first step in the analysis, the line spectra of each case are calculated according to Equation~\ref{eq:spec_line}.
The required  gradients are calculated with the same central (interior) and one-sided (boundaries) scheme \citep{Fornberg1988} as for the HIT case.

\begin{figure}[h]
\centering
\includegraphics[width=0.94\textwidth]{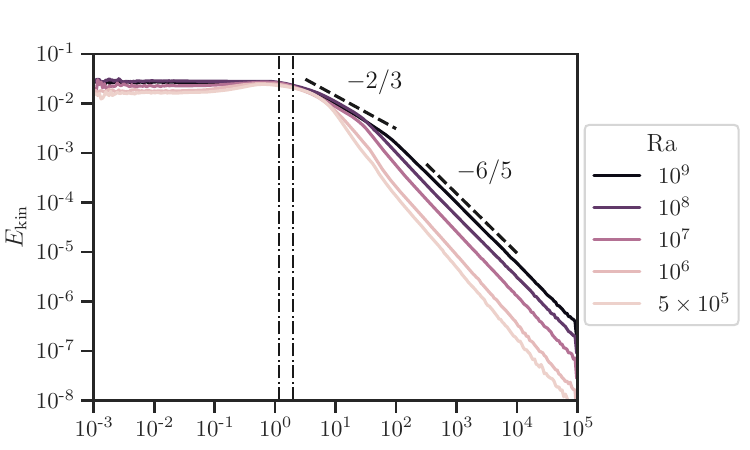}
\caption{Curvature-based energy spectra for a $\mathrm{Ra}$ variation ranging from $5\times10^5$ to $10^9$. Dash-dotted lines indicate the curvature values corresponding to the radii of the circumscribing and inscribing spheres of the cubic domain.}\label{fig:spec_gath}
\end{figure}

The resulting line spectra are plotted in Figure~\ref{fig:spec_gath}, where the scales associated with the domain size are marked by two dash-dotted lines referring to the curvatures of the spheres circumscribing and inscribing the cubic domain.
The comparison between the cases reveals that they all exhibit a similar maximum energy of $E_\mathrm{kin}\approx0.03$ for curvatures just above those associated with the domain size.
For smaller curvatures, they all display a constant behaviour with increasing $E_\mathrm{kin}$ values with increasing Rayleigh number.
For curvatures associated with the domain size and higher values, all cases pass the $-2/3$ slope of the inertial range before showing the bend towards a steeper slope with the onset of viscous effects.
This turning point is  \textendash\ as expected \textendash\ shifted towards larger curvatures for the more turbulent, high $\mathrm{Ra}$ cases, leading to higher kinetic energies contained in the larger curvatures of these cases. 

\begin{figure}[h]
\centering
\includegraphics[width=0.94\textwidth]{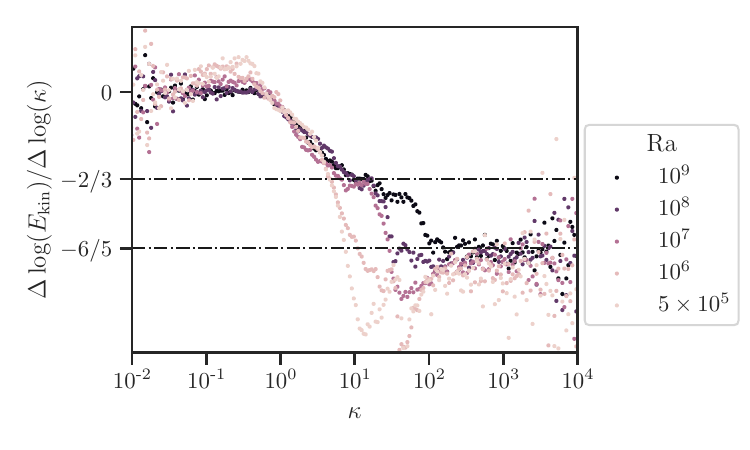}
\caption{Logarithmic slopes of the curvature-based spectra shown in Figure~\ref{fig:spec_gath}.}\label{fig:spec_gath_slopes}
\end{figure}

To analyse this in more detail, Figure~\ref{fig:spec_gath_slopes} displays the logarithmic slopes for adjacent $\kappa$ value pairs, in a comparable manner to a similar investigation of the second-order structure function by \citet{Barta2025}. It shows a significant transition from the lowest Rayleigh numbers $5\times10^5$ and $10^6$, which just pass the slope of $-2/3$ at $\kappa\approx10^1$, to the higher Rayleigh numbers, which exhibit a range starting at $\kappa\approx10^1$ in which the slopes stays at the value for the inertial range.
It is further observable that this inertial range slightly shifts to higher curvatures with increasing $\mathrm{Ra}$.

For even higher curvatures, all cases develop steeper slopes, which converge towards a value of $-6/5$ especially for the high Rayleigh number cases.
This is remarkable as this slope is associated with the Bolgiano-Obukhov scaling regime which expresses the influence of buoyancy onto the flow. This scaling is, however, expected to be prevalent at large structure sizes. While \citet{Alam2019}  show that this scaling can also be derived for small scales in stably-stratified turbulence, it is still unclear how this might be transferred to the present curvature-based investigation of RBC and more research is needed to complete understand the occurrence of the $-6/5$ slope at high curvatures.

\begin{figure}[h]
\centering
\includegraphics[width=0.95\textwidth]{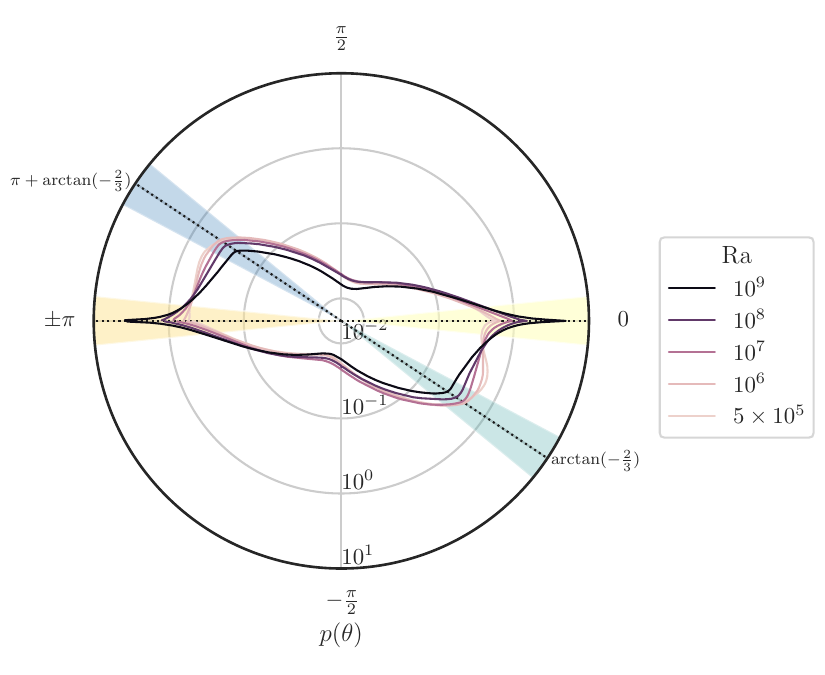}
\caption{Polar density distributions of the angle $\theta$ of the time evolution vector $\boldsymbol{U}^*$ for a $\mathrm{Ra}$ variation ranging from $5\times10^5$ to $10^9$. The angles corresponding to the $-2/3$ slope are marked by a dotted line. The coloured sectors refer to the regions for which, conditional sampling is executed for the back-projections displayed in Figure~\ref{fig:backproj}.}\label{fig:slopes_gath}
\end{figure}

In order to gain a deeper insight into the time evolution of the turbulent structures of the different flow cases, the next step of the analysis is to consider the time evolution vector $\boldsymbol{U}^*$ and to derive the respective polar density distributions of its slope angles $\theta$ with reference to Equations~\ref{eq:time_vec} to \ref{eq:dens_angle}.
For the sake of completeness, the streamline visualisations of $\boldsymbol{U}^*$ within the $E_\mathrm{kin}$-$\kappa$ plane are recorded in \ref{app:streamlines}.
The respective polar density distributions $p$ of the angle $\theta$ of the different Rayleigh numbers are collected in Figure~\ref{fig:slopes_gath}.
Starting from the most turbulent case $\mathrm{Ra}=10^9$, two pronounced modal values at $0$ and $\pm\pi$ can be observed, as was the case for HIT.
However, in contrast to HIT, the shoulders at the angles associated with the $-2/3$ slope are significantly more pronounced, which is emphasised by concave sections of $p(\theta)$.

Moving to smaller Rayleigh numbers, we are able to observe multiple effects.
First, the the density decreases for the peaks on the horizontal, while it increases for the $-2/3$ slope, even forming secondary peaks.
The second one is that the sections between the horizontal and the $-2/3$ slope is completely concave for $\mathrm{Ra}\geq10^7$, while for smaller $\mathrm{Ra}$, the secondary peaks form more towards the horizontals and create convex sections.
These secondary peaks of $\mathrm{Ra}\leq10^6$ even reach similar levels to the peak in the horizontal.
This supports the findings of a significant change of the behaviour of the $E_\mathrm{kin}(\kappa)$ slopes of the cases between $\mathrm{Ra}=10^6$ and $10^7$. Here, this change is expressed by the formation of distinct shoulders occurring exactly at the angles $\theta$ representing the $-2/3$ slope.

As discussed for the HIT case, the horizontal peaks represent perfect inertial behaviour of a fluid parcel.
Therefore, the $-2/3$ slope should not be viewed as a representation of the inertial range in the context of the time evolution vector $\boldsymbol{U}^*$.
Instead, we associate it with cascading behaviour as it shows a simultaneous increase in curvature and decrease in kinetic energy, and vice versa.
Two things are noteworthy in this context.
First, that the predicted $-2/3$ slope is not only statistically prevalent within the energy spectrum, but also plays a significant role in the time evolution experienced by a fluid parcel.
Second, that the upward and downward cascading parts of the density distribution are largely rotationally symmetric, suggesting some sort of symmetry of the respective processes within the flow.

In terms of the change in flow regime, the range of Rayleigh numbers studied lies in the region of the transition between soft and hard turbulence described by \citet{Castaing1989} and the transition between boundary layer dominated and bulk dominated thermal dissipation described by the Grossmann-Lohse-theory \citep{Grossmann2000, Stevens2013}.
It is therefore affirming to observe a clear footprint of a transition, in the form of a convex-concave change of the section between the horizontal and the $-2/3$ slope, associated with the inversion of the evolution of how pronounced the respective densities are.
\\

\begin{figure}[p]
    \begin{center}
        \begin{tabular}{l l l }
            & \multicolumn{1}{c}{\scriptsize $\mathrm{Ra}=10^6$} & \multicolumn{1}{c}{\scriptsize $\mathrm{Ra}=10^9$} \\ 
            
            {\scriptsize LSC} & \hspace{4em} $\circlearrowleft$ \hspace{6em} $\circlearrowright$ & \hspace{2.5em} $\circlearrowright$ \hspace{6em} $\circlearrowleft$\\
            
            \rotatebox{90}{\scriptsize \shortstack{pseudo \\ shadowgraph}} &  
            \includegraphics[trim={5 250 65 20},clip,scale=0.5]{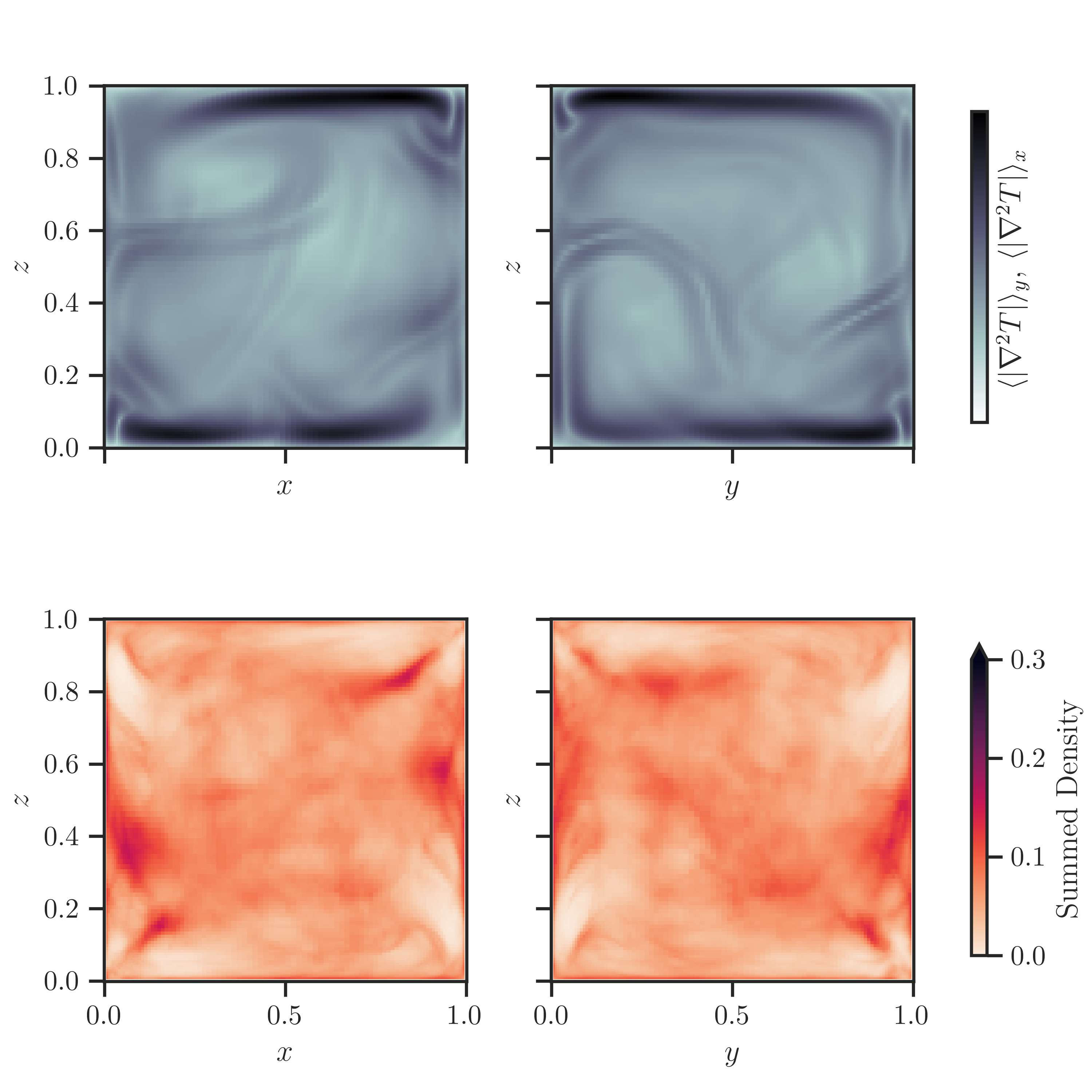} &  
            \includegraphics[trim={32 250 3 20},clip,scale=0.5]{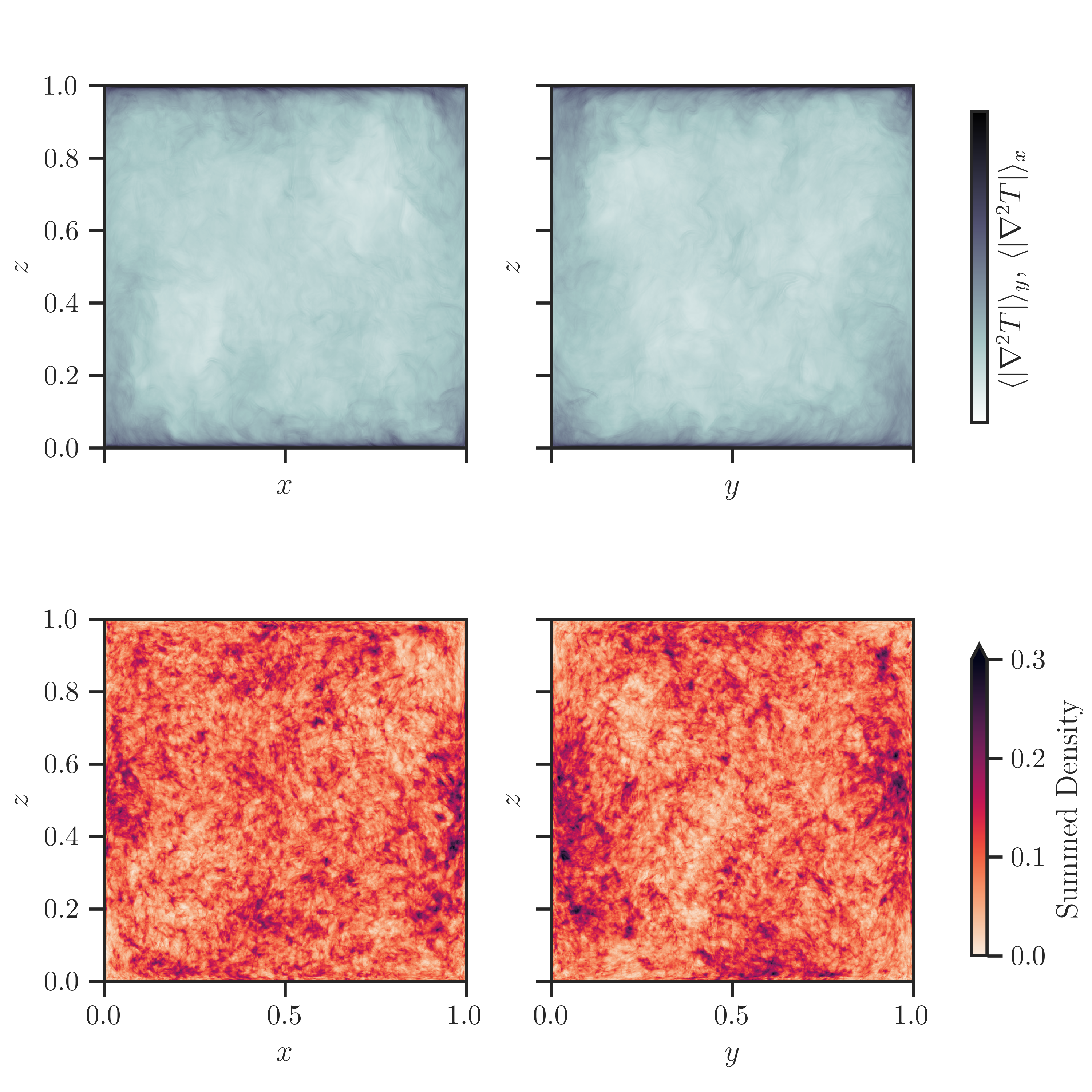} \\  

            \rotatebox{90}{\scriptsize \shortstack{curving \\ $\theta \approx 0$}} &  
            \includegraphics[trim={5 40 65 230},clip,scale=0.5]{fig_10_1.png} &  
            \includegraphics[trim={32 40 3 230},clip,scale=0.5]{fig_10_2.png} \\
 
            \rotatebox{90}{\scriptsize \shortstack{straightening \\ $\theta \approx \pi $}} &  
            \includegraphics[trim={5 40 65 230},clip,scale=0.5]{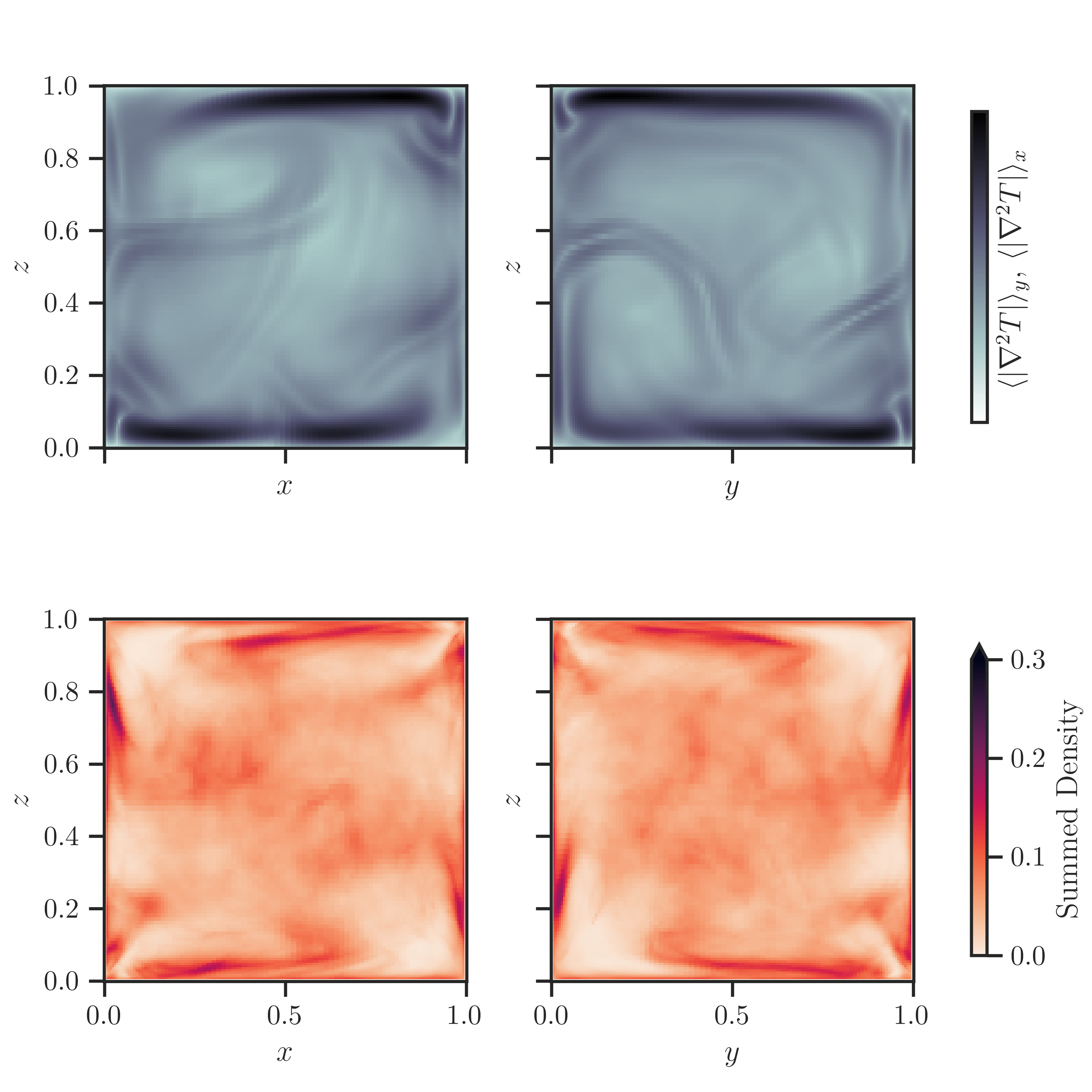} &  
            \includegraphics[trim={32 40 3 230},clip,scale=0.5]{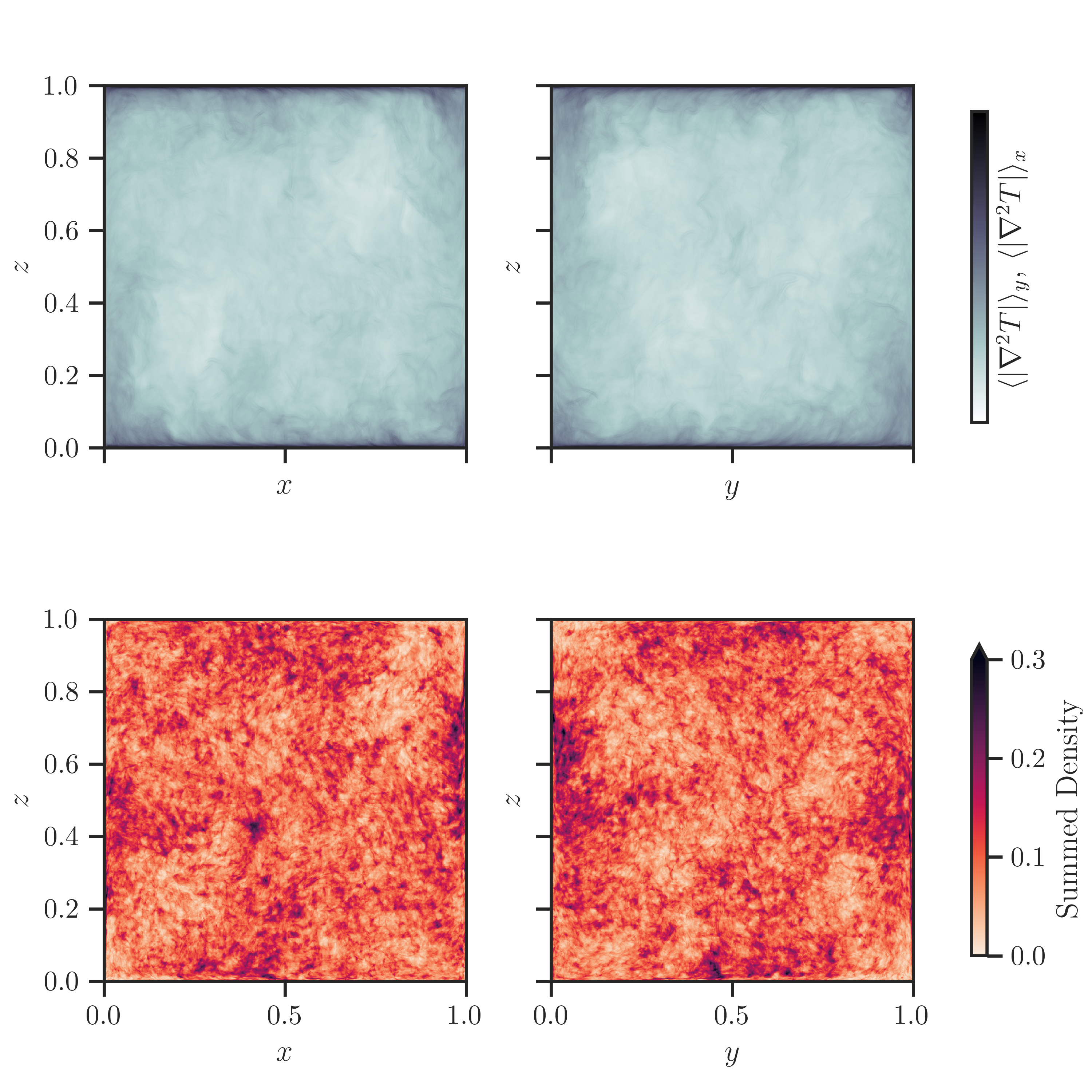} \\ 
 
            \rotatebox{90}{\scriptsize \shortstack{cascade down \\ $\theta \approx \arctan(\text{-}\frac{2}{3})$}} &  
            \includegraphics[trim={5 40 65 230},clip,scale=0.5]{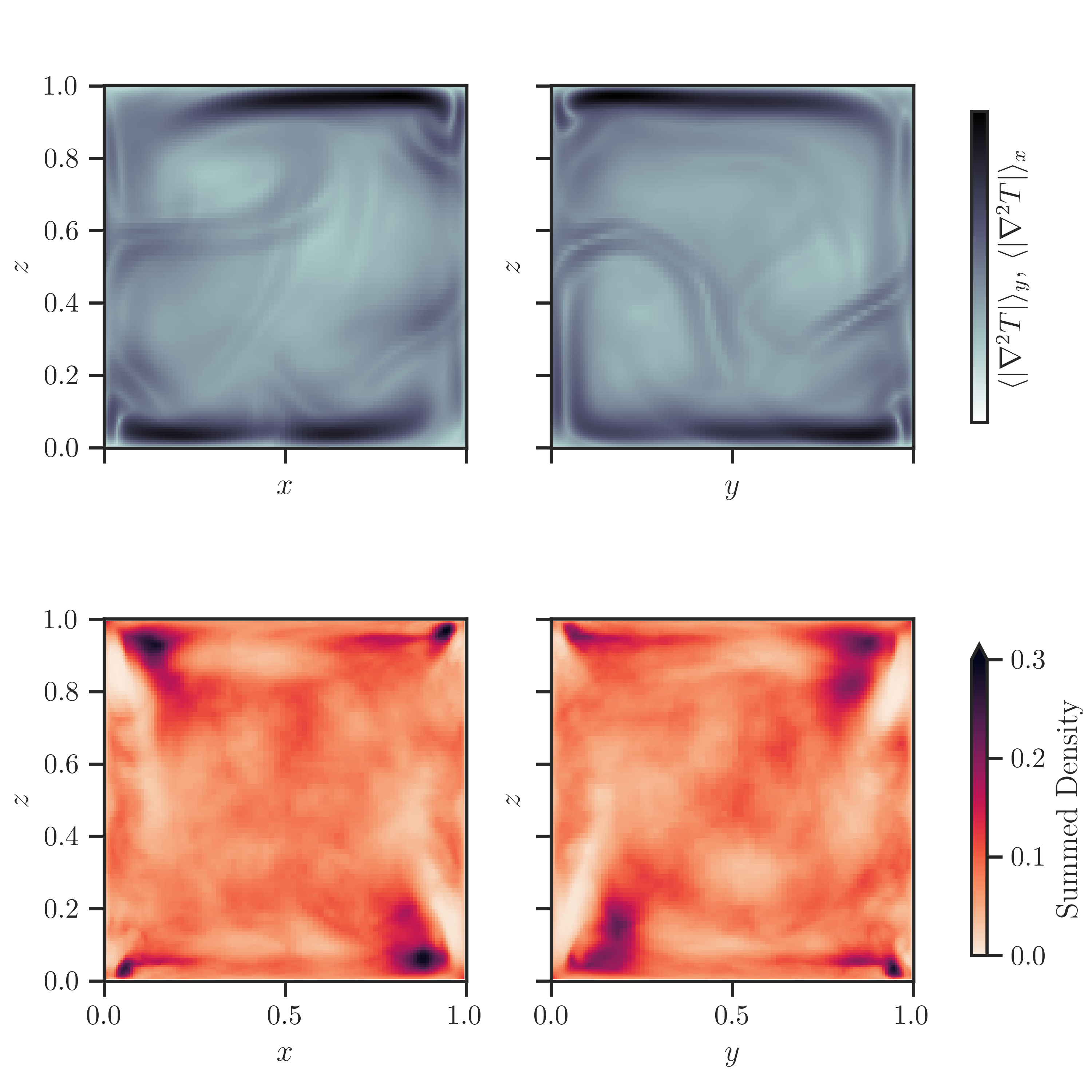} &  
            \includegraphics[trim={32 40 3 230},clip,scale=0.5]{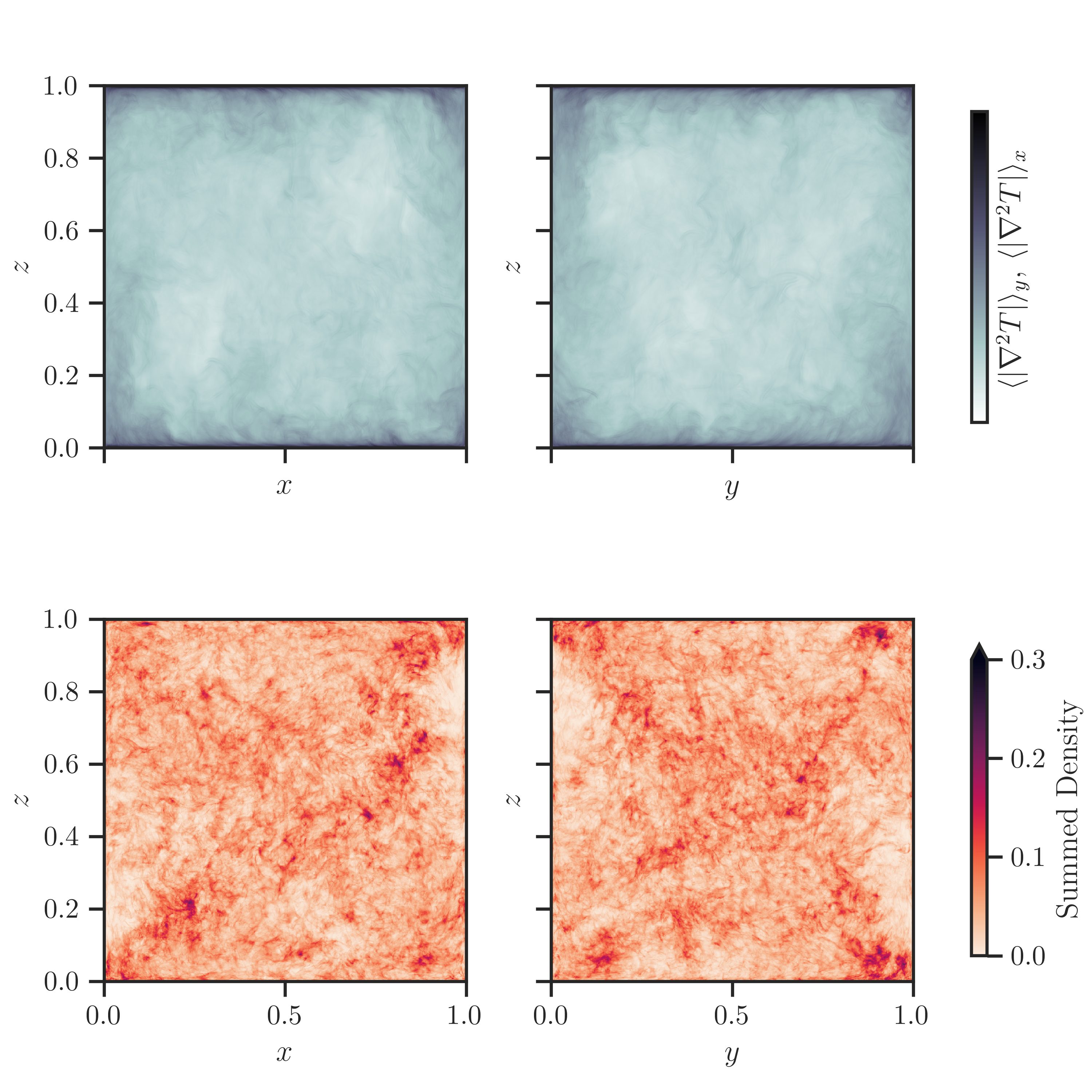} \\

            \rotatebox{90}{\scriptsize \shortstack{cascade up \\ $\theta \approx \pi + \arctan(\text{-}\frac{2}{3}) $}} &  
            \includegraphics[trim={5 10 65 230},clip,scale=0.5]{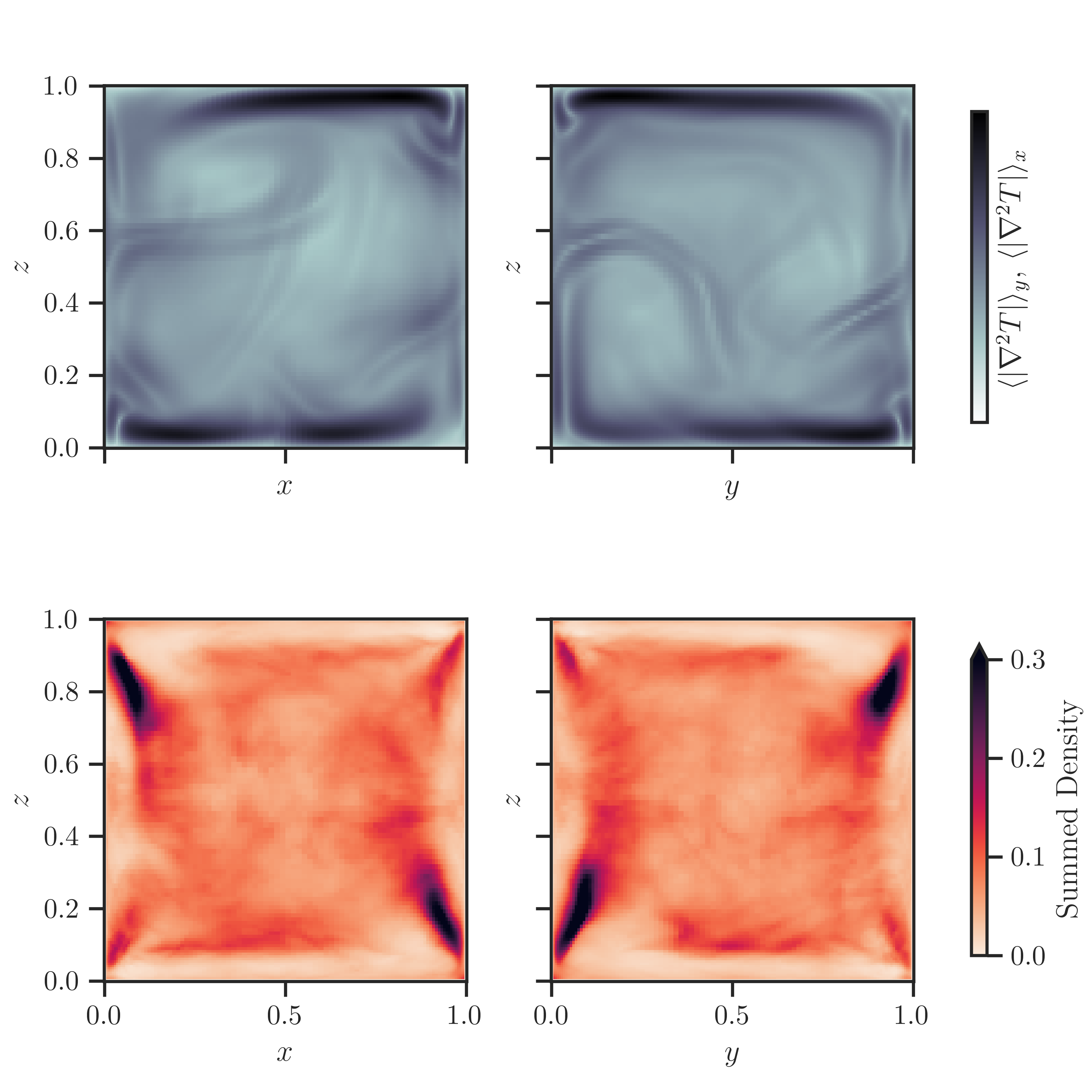} &  
            \includegraphics[trim={32 10 3 230},clip,scale=0.5]{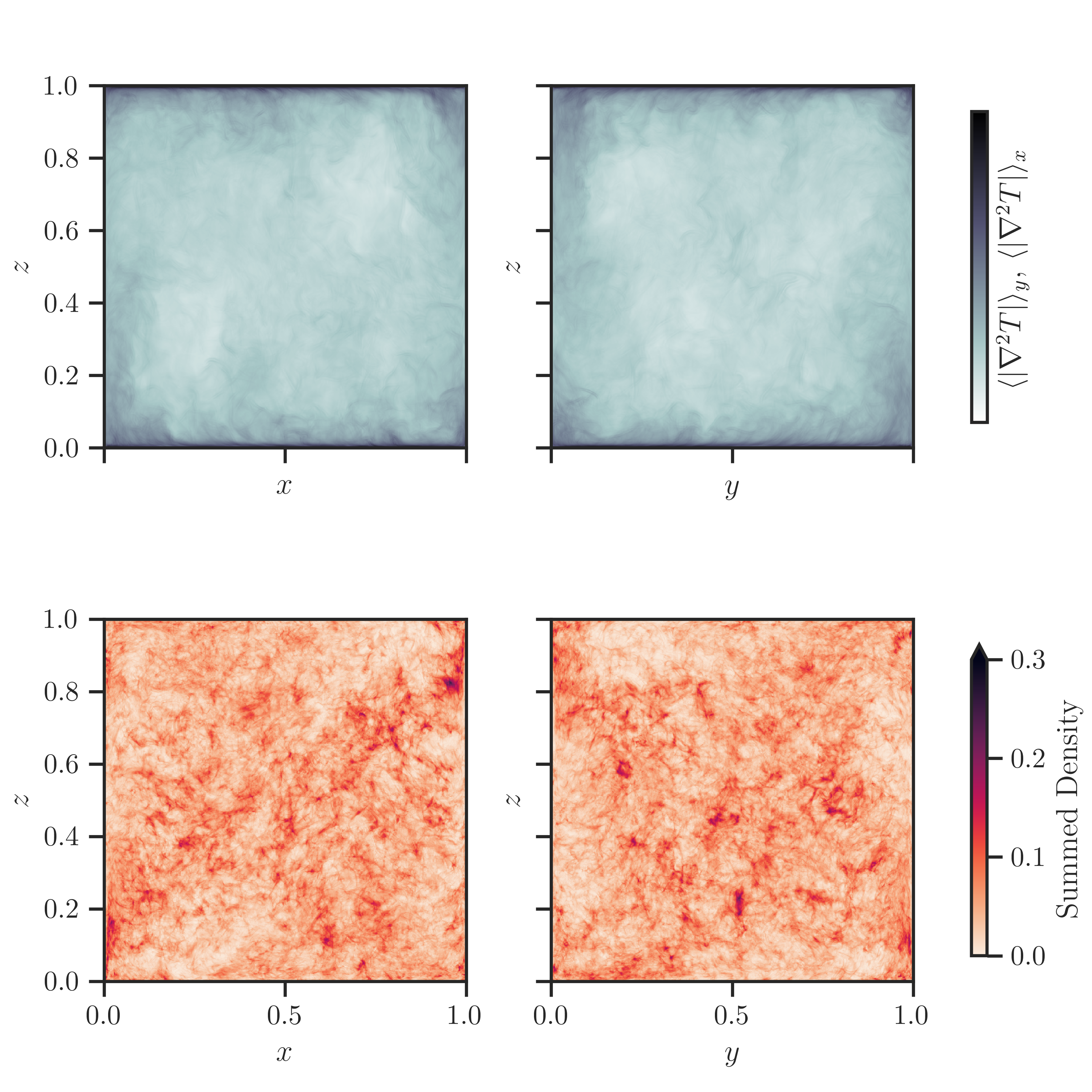} \\

        \end{tabular}

        \caption{Comparison of the back-projections of the $\theta$ ranges marked in Figure~\ref{fig:slopes_gath} into physical space for the Rayleigh numbers $10^6$ (left) and $10^9$ (right). The first row comprises instantaneous pseudo shadowgraphs ($\langle | \nabla^2 T | \rangle$), with an arrow indicating the rotation of the diagonal LSC in this section. For each $\mathrm{Ra}$, two horizontal viewing directions $y$-parallel (left) and $x$-parallel (right) are presented. Below, the incidence densities for the respective ranges of $\theta$ are displayed by integrating over the viewing direction and the respective time interval of the data set. }
        \label{fig:backproj}
    \end{center}
\end{figure}

In order to better understand the effects observed for the density distribution $p(\theta)$, we take advantage of another benefit of the persistent link to the original data, namely the possibility of conditional sampling and back-projection.
Specifically, we define regions of interest of $\pm 0.1$ radians around the salient angles of $p(\theta)$, which are marked in Figure~\ref{fig:slopes_gath} as follows: Yellow ($\theta\approx0$) indicates the inertial \textit{curving} section opposite to the \textit{straightening} section marked in orange ($\theta\approx\pm\pi$). Similarly, the green section ($\theta\approx\arctan(-\frac{2}{3})$) marks \textit{downward cascading} behaviour, while the blue section ($\theta\approx\pi+\arctan(-\frac{2}{3})$) represents the opposite \textit{upward cascading} behaviour.
These segments were then used as conditions to select the samples of the data sets to be back-projected into the physical domain.
To give an impression of where these samples are located, their summed densities, i.e. viewing direction-integrated and time-averaged locations, are plotted in Figure~\ref{fig:backproj} for the cases of $\mathrm{Ra}=10^6$ (left) and $\mathrm{Ra}=10^9$ (right). To account for the three-dimensional nature of the flow, both horizontal viewing directions, $y$ (left) and $x$ (right), are displayed for each case.
For orientation, the first row provides instantaneous pseudo-shadowgraphs, i.e. fields of $\langle | \nabla^2 T | \rangle_y$ and $\langle | \nabla^2 T | \rangle_x$, as they also integrate along the viewing direction and provide information about the diagonal orientation of the LSC.

Examining the back-projections for the $\mathrm{Ra}=10^6$ case (left side of Figure~\ref{fig:backproj}), it becomes apparent that the purely inertial processes of curving and straightening occur somewhat less frequently than their cascading counterparts.
Pure curving appears to be relatively uniformly distributed, but with prominent maxima at the side walls, where plumes begin to curve before impacting the opposing horizontal plates. Minima for curving are visible for regions where the pure straightening has its maxima, namely the impact regions of the top and bottom plates as well as regions at the side walls associated with plume ejection.
The minimum regions of the straightening process are mainly occupied by the two cascading processes.
For downward cascading, i.e. curving with decreasing kinetic energy, the maxima can be found in the vicinity of the horizontal plates, close to the edges where the plumes detach. However, the minima for this process are located immediately adjacent to the maxima, at the side walls, where the detached plumes straighten.
This is also where the maxima for the upward cascade are located, as the fluid is typically accelerated by buoyancy in these regions.
In addition to the regions of these maxima, the samples with upward cascading behaviour tend to be found in the bulk region.

On the other side, the $\mathrm{Ra}=10^9$ case exhibits density maps with much finer structures,  which is due to the short averaging time interval of this case. Nevertheless, the regions where certain processes are concentrated, are already observable.
Consistent with the results of the densities of $p(\theta)$ (see Figure~\ref{fig:slopes_gath}), the two purely inertial processes in this case have higher incidences compared to with the lower $\mathrm{Ra}$.
The maximum regions for the curving and straightening behaviour appear very similar and are located in the vicinity of all the walls, including the horizontal plates. 
Due to the finer structures, it is harder to observe whether the adjacent positioning of downward and upward cascading persists for the high Rayleigh number.
Yet, the detaching regions of the plumes still exhibit slightly higher densities for downward cascading behaviour.

\begin{figure}[bh!]
\centering
\includegraphics[width=0.66\textwidth]{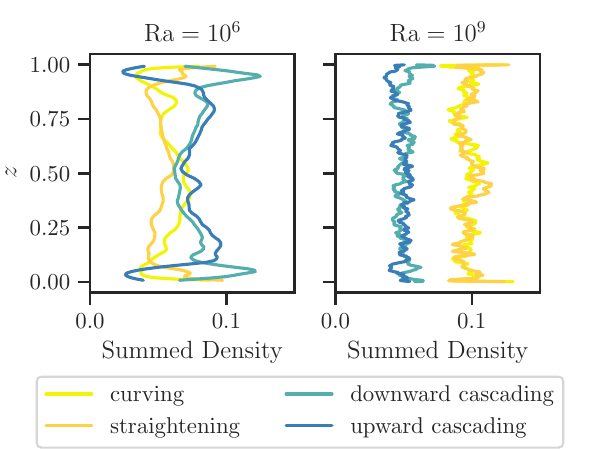}
\caption{$z$-profiles of the summed densities of the various behaviours curving, straightening, downward cascading, and upward cascading shown in Figure~\ref{fig:backproj}.}\label{fig:z_profiles}
\end{figure}

To provide a more quantitative view, the $z$-profiles of the discussed $\theta$-sections are displayed in Figure~\ref{fig:z_profiles} for both cases.
They reflect the already discussed trend of higher fractions of the the pure curving and straightening behaviour for the higher Rayleigh number.
Further, the Figure shows a transition from spatially concentrated occurrences at $\mathrm{Ra}=10^6$, e.g. for the downward cascading with peaks close to the top and bottom plates, to significantly more uniform distributions for the case of $\mathrm{Ra}=10^9$.
With regard to the earlier described transition of the flow regime, the aspect of the change in vertical distribution of the downward cascading behaviour correlates well with the shift of the dominance of thermal dissipation from the boundary layer to the bulk as described by the Grossmann-Lohse-theory.

\subsection{PTV data}\label{sec:ptv}

\begin{figure}[h!]
    \begin{center}
        \begin{tabular}{c c }
            a) & b) \\
            \includegraphics[trim={55 75 30 125},clip,width=0.5\textwidth]{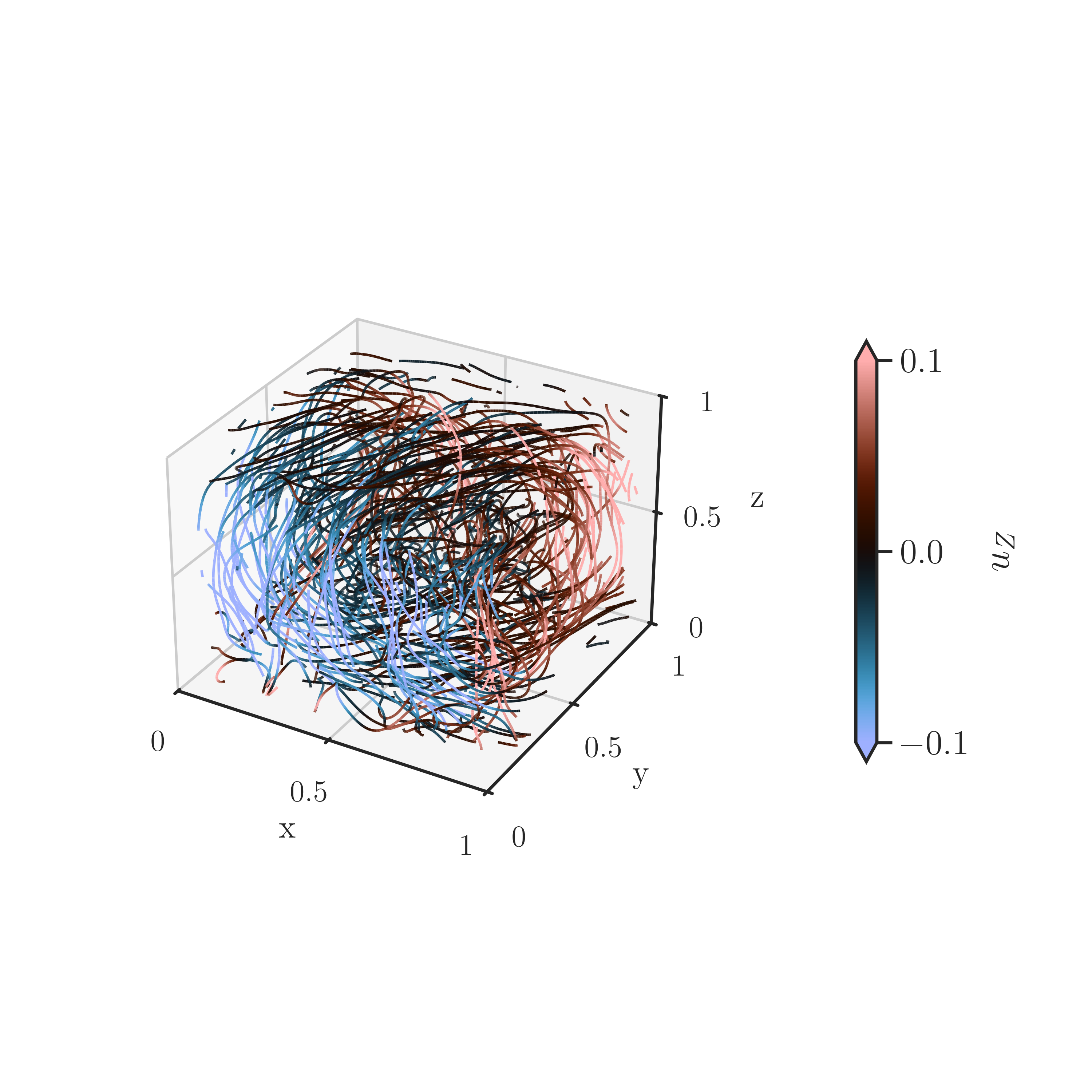} &  
            \includegraphics[trim={0 0 0 0},clip,width=0.5\textwidth]{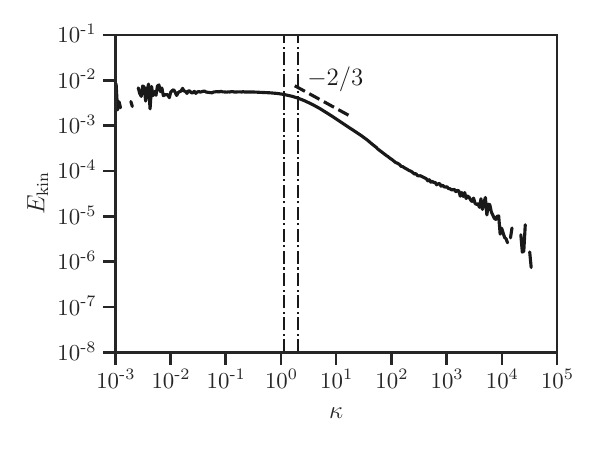} \\
        \end{tabular}
        \caption{a) Visualisation of 500 exemplary particle tracks from the LPT data set. b) Curvature-based energy spectrum of the LPT data set at $\mathrm{Ra} = 2.5 \times 10^9$ and $\mathrm{Pr} = 7$.}
        \label{fig:tracks}
    \end{center}
\end{figure}

To demonstrate the feasibility of applying the curvature-based analysis framework to measurement data, we introduce a Lagrangian particle tracking (LPT) dataset with $\mathrm{Ra}=2.5 \times 10^9$ and $\mathrm{Pr}=7$. It is part of the study by \citet{Barta2024} and was acquired using of the proPTV measurement framework \citep{Barta2024a}.
Figure~\ref{fig:tracks} displays $500$ exemplary tracks out of the total of $169428$ tracks used for the analysis. These tracks fill an interval of approximately $253$ free-fall time units and yield approximately $2.5 \times 10^7$ samples with a time resolution of $\delta t \approx 0.042$.
A Savitzky-Golay filter \citep{Savitzky1964} with a polynomial degree of $3$ and a window size of $30$ time steps has been applied to eliminate measurement noise. 

The first step of the analysis is again the calculation of the spectrum $E_\mathrm{kin}(\kappa)$ according to Equation~\ref{eq:spec_line}, which is displayed in Figure~\ref{fig:tracks}b).
For this, we assume that the tracer particles are uniformly distributed, meaning that the volume weighting was not applied for this case.
As for the DNS cases with a lower $\mathrm{Pr}$, the spectrum shows a constant behaviour in the range of the smallest curvatures up to the curvatures associated with the domain size. 
From there on, a decrease in the kinetic energy with the $-2/3$ slope is observed, which becomes steeper in the region $10 \lesssim \kappa \lesssim 30$.
At $\kappa \approx 100$, the spectrum bends towards less steep slopes again.
As there is no viable physical explanation for this, we attribute this behaviour to the inertia of the tracer particles, which also prevents measurements for very high curvatures ($\kappa \gtrsim 10^4$).

\begin{figure}[h]
\centering
\includegraphics[width=0.95\textwidth]{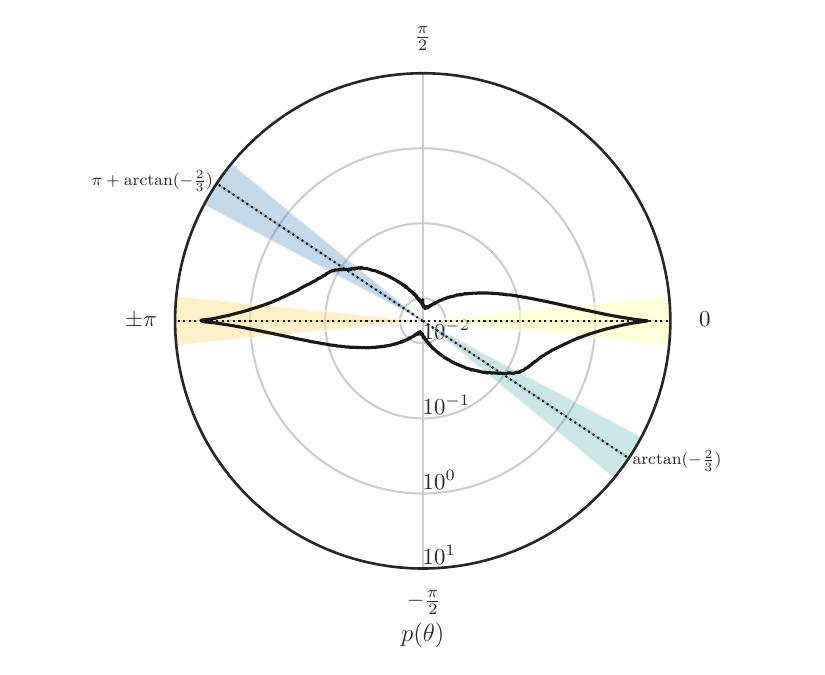}
\caption{Polar density distribution of the angle $\theta$ of the time evolution vector $\boldsymbol{U}^*$ for the LPT data set at $\mathrm{Ra} = 2.5 \times 10^9$ and $\mathrm{Pr} = 7$.}\label{fig:tracks_slopes}
\end{figure}

The next key point of the curvature-based analysis framework is the density distribution $p(\theta)$ for the time evolution vector $\boldsymbol{U}^*$ displayed in Figure~\ref{fig:tracks_slopes}.
As with the curvature spectrum, this polar density distribution plot of the LPT dataset displays features also seen for the DNS dataset of similar Rayleigh number and lower Prandtl number.
Specifically, the dominant modal values at angles $0$ and $\pm \pi$ with a concave segment extending up to the angles for the $-2/3$ slope, where they form a pronounced shoulder, are replicated.
Also taking into account possible time resolution effects discussed in \ref{app:stat_sensitivity}, the only qualitative deviation from the DNS results is that the locations of the shoulders do not exactly match the angles of the $-2/3$ slope.
As these locations are slightly shifted towards the horizontal, which represents purely inertial behaviour, we also attribute this to the inertial effects of the tracer particles.

\section{Conclusion and Outlook}

To avoid the spurious effects of an FFT in bounded domains, we introduced an analysis framework based on curvature as a measure of the flow structure size.
Its application to homogeneous isotropic turbulence showed that curvature-based energy spectra are able to cover the main characteristics of the flow, such as the slope for the inertial range or the onset of viscous effects, while not being completely congruent with the results provided by classical energy spectra.

We further showed that the persistent link to the raw data is an advantage of this method, which can be exploited to learn about the temporal evolution fluid parcels experience as well as \textendash\ by means of back-projection \textendash\ the locations where certain behaviours predominantly occur.

These two features proved to be insightful for the application of the framework to a variety of cases of cubic Rayleigh-Bénard convection, with a Prandtl number of $0.7$ and Rayleigh numbers ranging from $5\times10^5$ to $10^9$.
Within this range, the transition from soft to hard turbulence is expected, as well as a switch from boundary layer-dominated to bulk-dominated thermal dissipation, as described by the Grossmann-Lohse-theory.
Using the presented framework, we were able to show the following characteristics of this transition:
The curvature-based energy spectra reflect the transition of the flow regime by exhibiting a distinct inertial range with a slope of $-2/3$ for $\mathrm{Ra}\geq10^7$ besides the expected effect of overall higher kinetic energies at higher curvatures towards larger Rayleigh numbers.
For the polar distribution of the time evolution vectors, this transition appears as a smooth one, as it is described for the transition between soft and hard turbulence.
Here, the smooth transition is characterised by an increase of the purely inertial curving and straightening behaviour of fluid parcels towards higher Rayleigh numbers with a corresponding decrease in the cascading behaviour.
Nonetheless, at the transition point between $\mathrm{Ra}=10^6$ and $10^7$ observed in the curvature-based energy spectra, the polar distributions start to exhibit distinct shoulders at angles associated to the $-2/3$ slope.
This means that this slope has not only a meaning in a purely statistical sense but also in the actual time development of individual fluid parcels.

Subsequently, conditional sampling and back-projection of the various behaviours into the physical domain were conducted for the Rayleigh numbers $10^6$ and $10^9$. The respective spatial distribution and especially the associated $z$-profiles showed that the cascading behaviour shifts to a less localised and therefore more bulk-centred occurrence during the transition of the flow regime.

Finally, we showed that it is possible to apply this framework to Lagrangian particle tracking data sets, which yielded consistent and physically plausible results.
This makes this method also appealing for the study of large-scale natural flows, which often have unclear boundaries and are difficult to fully resolve in numerical simulations.
Therefore, the observed deficiencies, caused by inertial effects of the tracer particles, need to be overcome to cover a wider range of curvatures.
One possible solution would be to combine Lagrangian particle tracking with emerging data assimilation methods using physics-informed neural networks. For these, \citet{Mommert2024} and \citet{Toscano2025} showed successful applications to thermal convective flows and \citet{Zhou2023} showed the possibility of extracting the inertial effects of tracer particles.

\section*{Acknowledgment}
The authors gratefully acknowledge the scientific support and HPC resources provided by the German Aerospace Center (DLR). The HPC system CARO is partially funded by "Ministry of Science and Culture of Lower Saxony" and "Federal Ministry for Economic Affairs and Climate Action".

The authors thank Theo Käufer and Christian Cierpka for the fruitful discussion on the role of the types of curvature definition and the applied time derivative.

\section*{Declaration of generative AI and AI-assisted technologies in the writing process}

During the preparation of this work the author(s) used DeepL in order to conduct language editing. After using this tool/service, the author(s) reviewed and edited the content as needed and take(s) full responsibility for the content of the publication.

\appendix
\section{Probability density of curvatures for an idealised circulation}\label{app:curv_distr_circ}

We consider a synthetic, stationary, circular flow centred around an origin, with stream lines effectively forming a set of concentric circles.
The radius of curvature of any one circle $r_i$ is equal to its distance from the origin $R$. 

Consequently, when sampling the curvatures at points uniformly and at random, the probability of the sample corresponding to a curvature radius between $r$ and $r+\mathrm{d}r$ is proportional to the area of the circular ring bounded by circles with radii $r$ and $r+\mathrm{d}r$, respectively:
\begin{align}
    P(r,r+\mathrm{d}r) \propto \pi (r+\mathrm{d}r)^2 - \pi r^2 = 2\pi r \mathrm{d}r + \mathcal{O}(\mathrm{d}r^2).\label{eq:p_radius}
\end{align}

Given the definition of the curvature $\kappa=1/r$,
\begin{align}
    \frac{\mathrm{d}r}{\mathrm{d}\kappa} = -\frac{1}{\kappa^2}\Rightarrow \mathrm{d}r = -\kappa^{-2}\; \mathrm{d}\kappa.
\end{align}
Substituting $r$ and $\mathrm{d}r$ in (\ref{eq:p_radius}) yields the probability density function for the curvature
\begin{align}
    f(\kappa) \propto 2\pi \kappa^{-3}.
\end{align}

This shows that the exponential decay of the relative incidences towards high curvatures for the synthetic flow representing a wave-number of $1$ shown in Figure~\ref{fig:gen_vort} is a purely geometric one.

\section{Considerations on temporal convergence}\label{app:stat_sensitivity}

The varying number of time frames $N_t$ considered naturally raises questions about the convergence of the analyses presented.

\begin{figure}[h!]
    \begin{center}
    \includegraphics[trim={0 0 0 0},clip,width=0.95\textwidth]{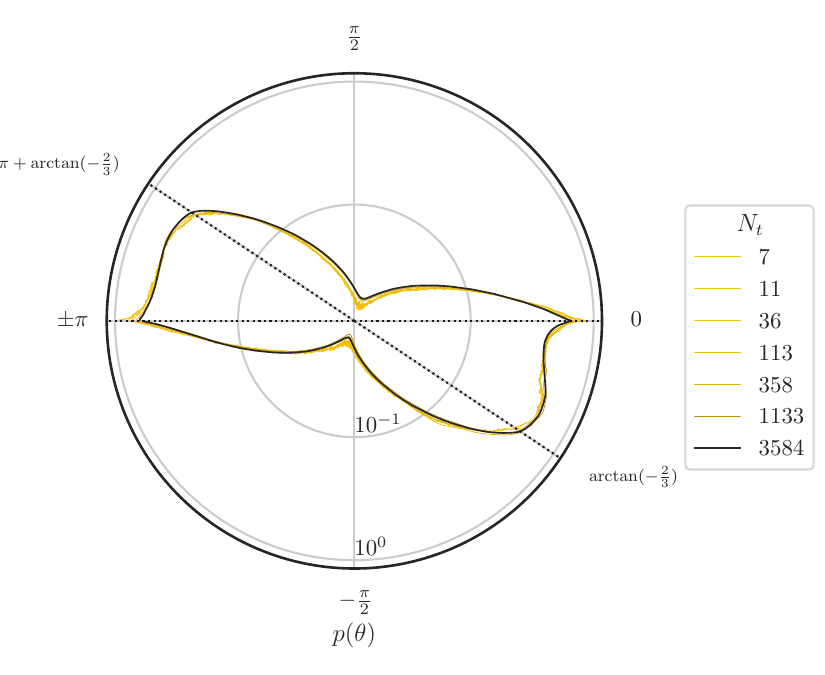}
    \caption{Polar density distribution of the angle $\theta$ of the time evolution vector $\boldsymbol{U}^*$ for $\mathrm{Ra} = 10^6$ and varying numbers of frames $N_t \in \{3584;1133;358;113;36;11;7\}$ considered.}\label{fig:dframe_1e6}
    \end{center}
\end{figure}

The first question concerns whether short time intervals $N_t \times \delta t$ affect the shape of the polar density distributions $p(\theta)$ of the angle of the time evolution vector $\boldsymbol{U}^*$.
To address this Figure~\ref{fig:dframe_1e6} plots $p(\theta)$ for a varying number of frames considered $N_t$.
It shows that even for the smallest number of frames number investigated, $N_t=7$, the deviations from the distribution for $N_t=3584$ (black) are minimal and lower $N_t$ are mainly characterised by more pronounced noise.
Therefore, the shape of $p(\theta)$ appears to be independent of the time interval considered.

\begin{figure}[h!]
    \begin{center}
    \includegraphics[trim={0 0 0 0},clip,width=0.95\textwidth]{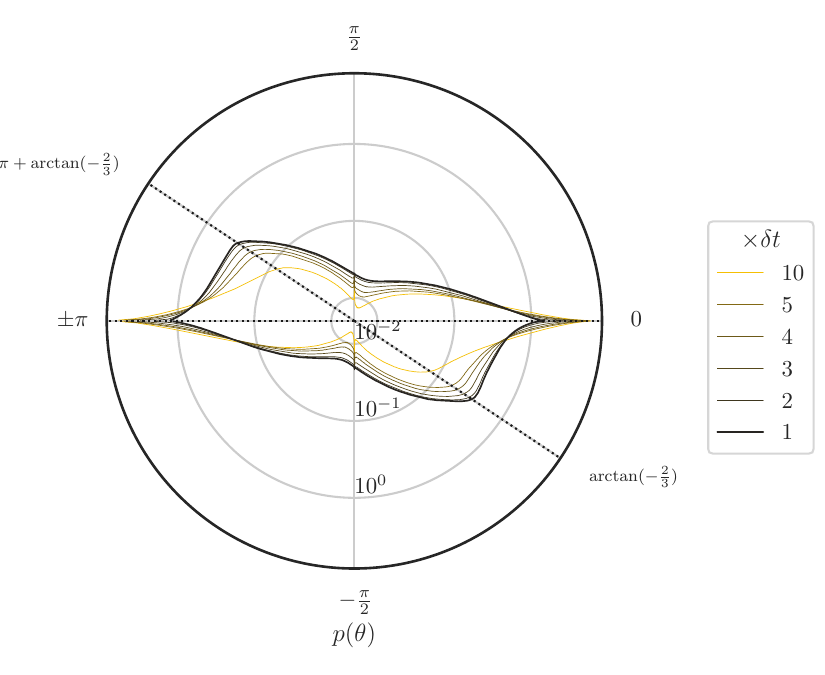} 
    \caption{Polar density distribution of the angle $\theta$ of the time evolution vector $\boldsymbol{U}^*$ for $\mathrm{Ra} = 10^8$ and varying frame separations $\delta t \in \{0.01;0.02;0.03;0.04;0.05;0.1\}$ considered.}\label{fig:dt_1e8}
    \end{center}
\end{figure}

To investigate the influence of the time resolution $\delta t$ of the provided frames, we regard the cases of $\mathrm{Ra} = 10^8$ and $\mathrm{Ra} = 10^9$ as they are the most demanding cases regarding the respective $\delta t$.
The corresponding variation in $\delta t$ for $\mathrm{Ra} = 10^8$ is shown in Figure~\ref{fig:dt_1e8}.
It reveals that a low time resolution results in more pronounced modal values at the angles of $0$ and $\pm \pi$ and less pronounced shoulders or secondary peaks at the angles of the $-2/3$ slope.
However, since $\delta t=0.02$ and $\delta t=0.01$ are almost congruent, we do not expect significant changes in the shape of $p(\theta)$ for higher time resolutions.

\begin{figure}[h!]
    \begin{center}
    \includegraphics[trim={0 0 0 0},clip,width=0.95\textwidth]{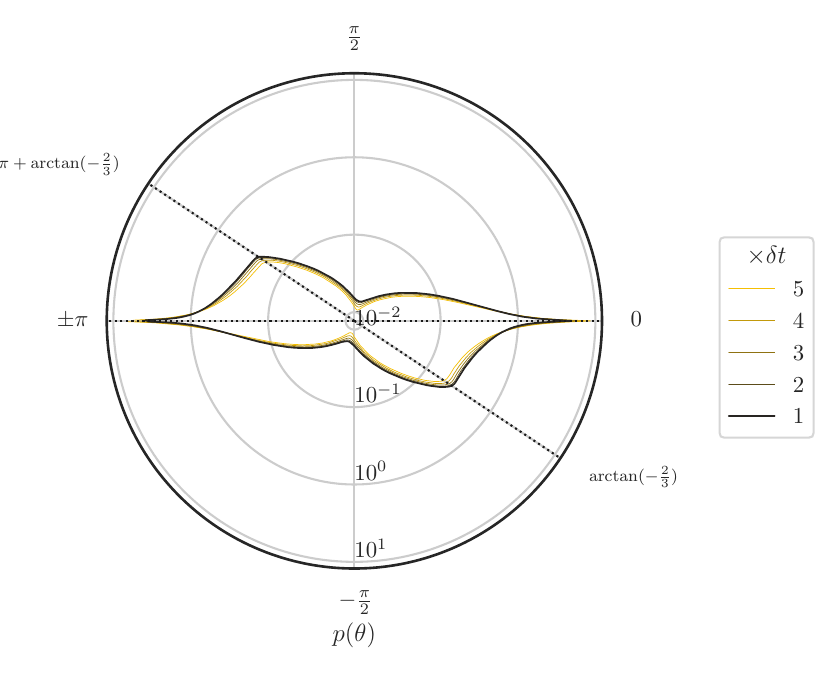} 
    \caption{Polar density distribution of the angle $\theta$ of the time evolution vector $\boldsymbol{U}^*$ for $\mathrm{Ra} = 10^9$ and varying frame separations $\delta t \in \{0.001;0.002;0.003;0.004;0.005\}$ considered.}\label{fig:dt_1e9}
    \end{center}
\end{figure}

Figure~\ref{fig:dt_1e9} shows the respective plot for $\mathrm{Ra} = 10^9$. It displays the same effects of low time resolution as for $\mathrm{Ra} = 10^8$. The used frame separation of $\delta t=0.001$ is also considered as sufficient, as its statistics are barely distinguishable from the one calculated with  $\delta t=0.002$.

\begin{figure}[h!]
    \begin{center}
        \begin{tabular}{c c c}
            a) & b) & c)\\
            \includegraphics[trim={0 0 0 0},clip,width=0.33\textwidth]{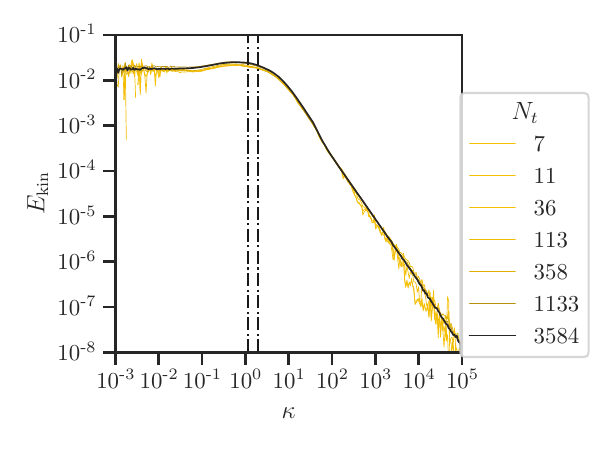} & 
            \includegraphics[trim={0 0 0 0},clip,width=0.33\textwidth]{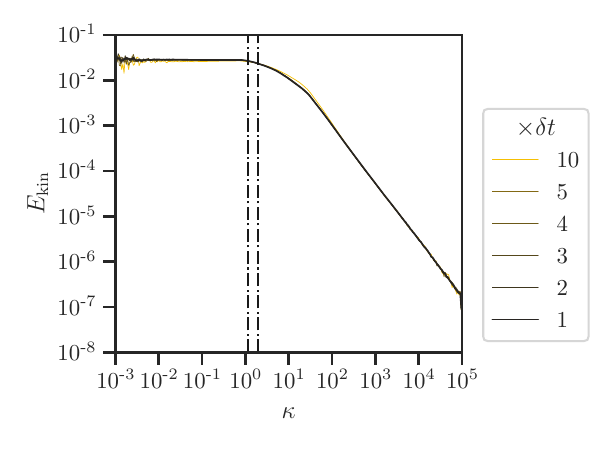} &
            \includegraphics[trim={0 0 0 0},clip,width=0.33\textwidth]{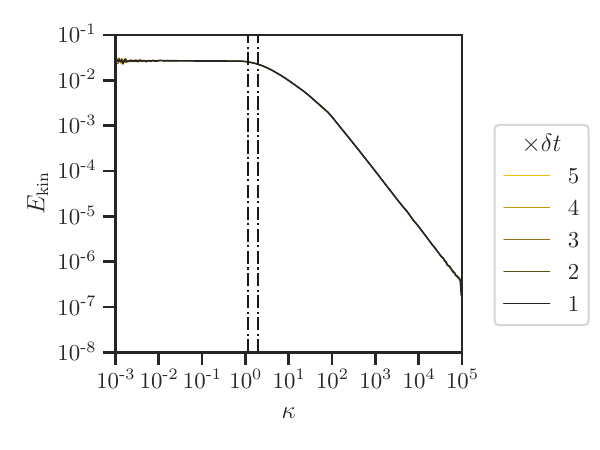} \\
        \end{tabular}
        \caption{Curvature-based energy spectra for the sensitivity studies discussed above: a) $N_t$ variation for $\mathrm{Ra} = 10^6$  b) $\delta t$ variation for $\mathrm{Ra} = 10^8$ c) $\delta t$ variation for $\mathrm{Ra} = 10^9$.}
        \label{fig:sensi_spectra}
    \end{center}
\end{figure}

For the sake of completeness, Figure~\ref{fig:sensi_spectra} displays the curvature-based energy spectra for the sensitivity studies discussed. As these do not rely on the additional time derivative introduced by $\boldsymbol{U}^*$, they are even more robust.
Therefore, the only significant deviations visible in these plots are those due to the noise of the inadequate statistics at very low sample counts.

\section{Streamline visualisations of $\boldsymbol{U}^*$}\label{app:streamlines}

\begin{figure}[p]
    \begin{center}
        \begin{tabular}{c c }
            a) & b)  \\
            \includegraphics[trim={0 0 0 0},clip, width=0.49\textwidth]{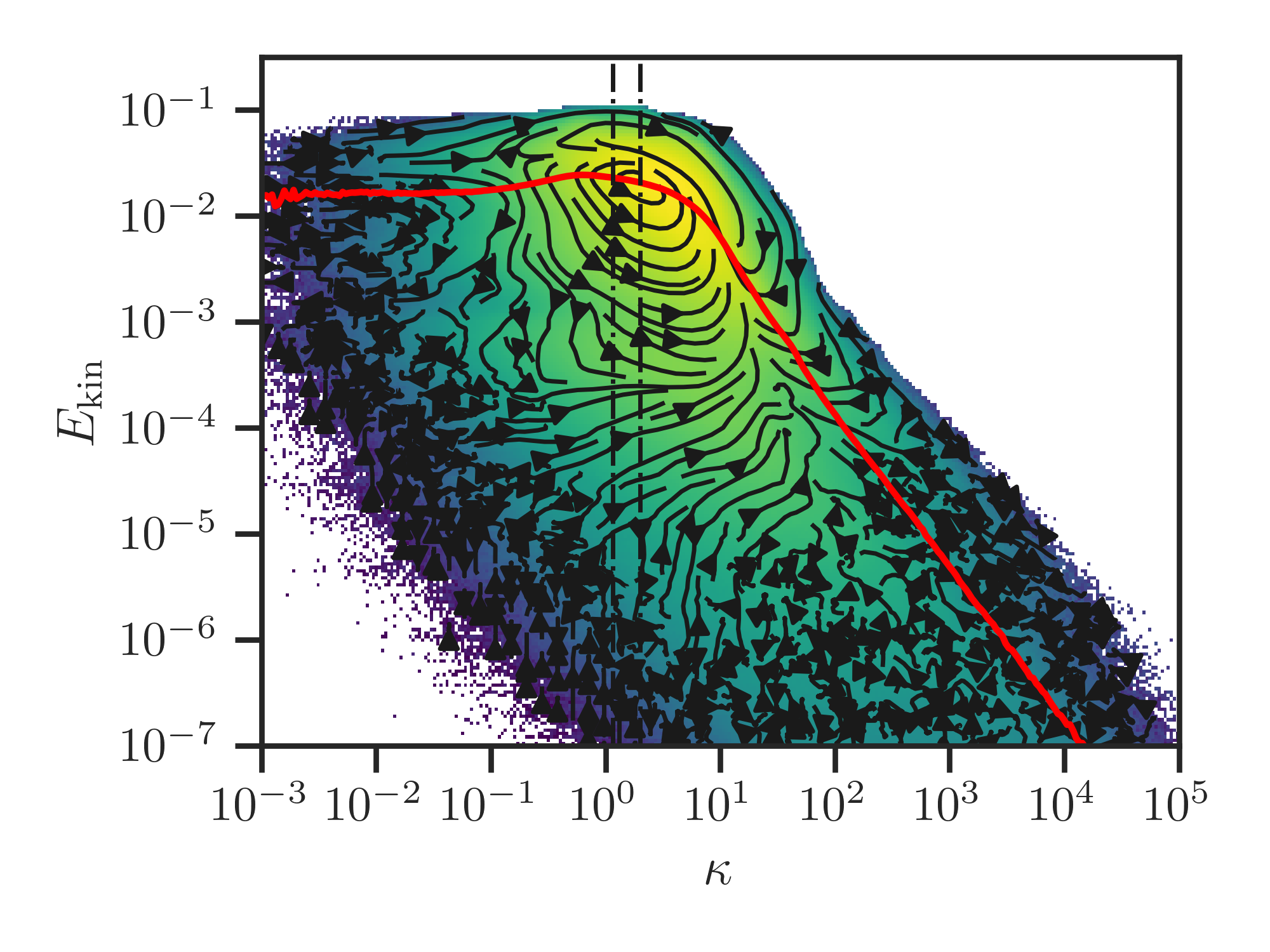} &
            \includegraphics[trim={0 0 0 0},clip, width=0.49\textwidth]{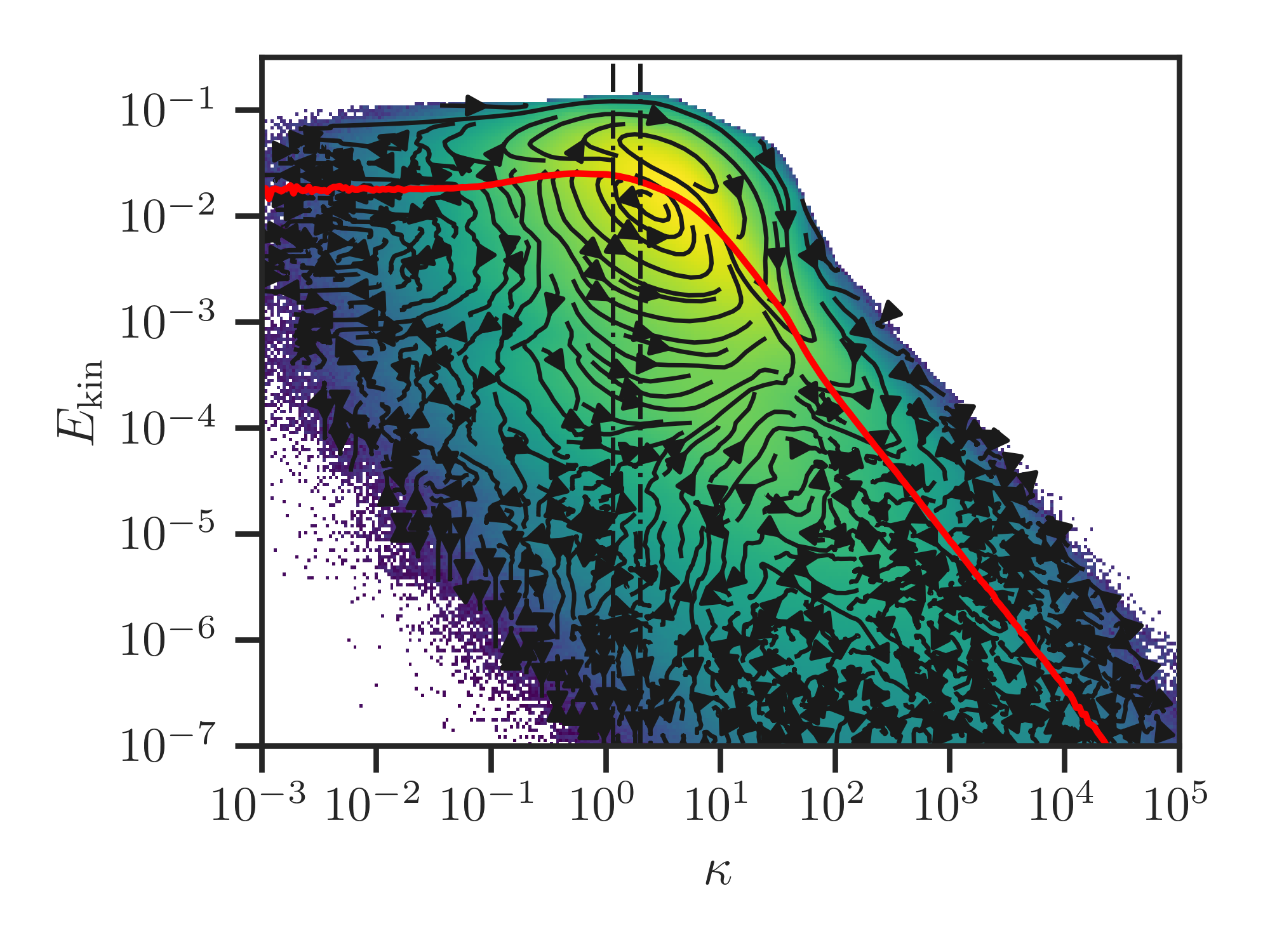} \\
        \end{tabular}
        \begin{tabular}{c c}
            c) & d) \\
            \includegraphics[trim={0 0 0 0},clip, width=0.49\textwidth]{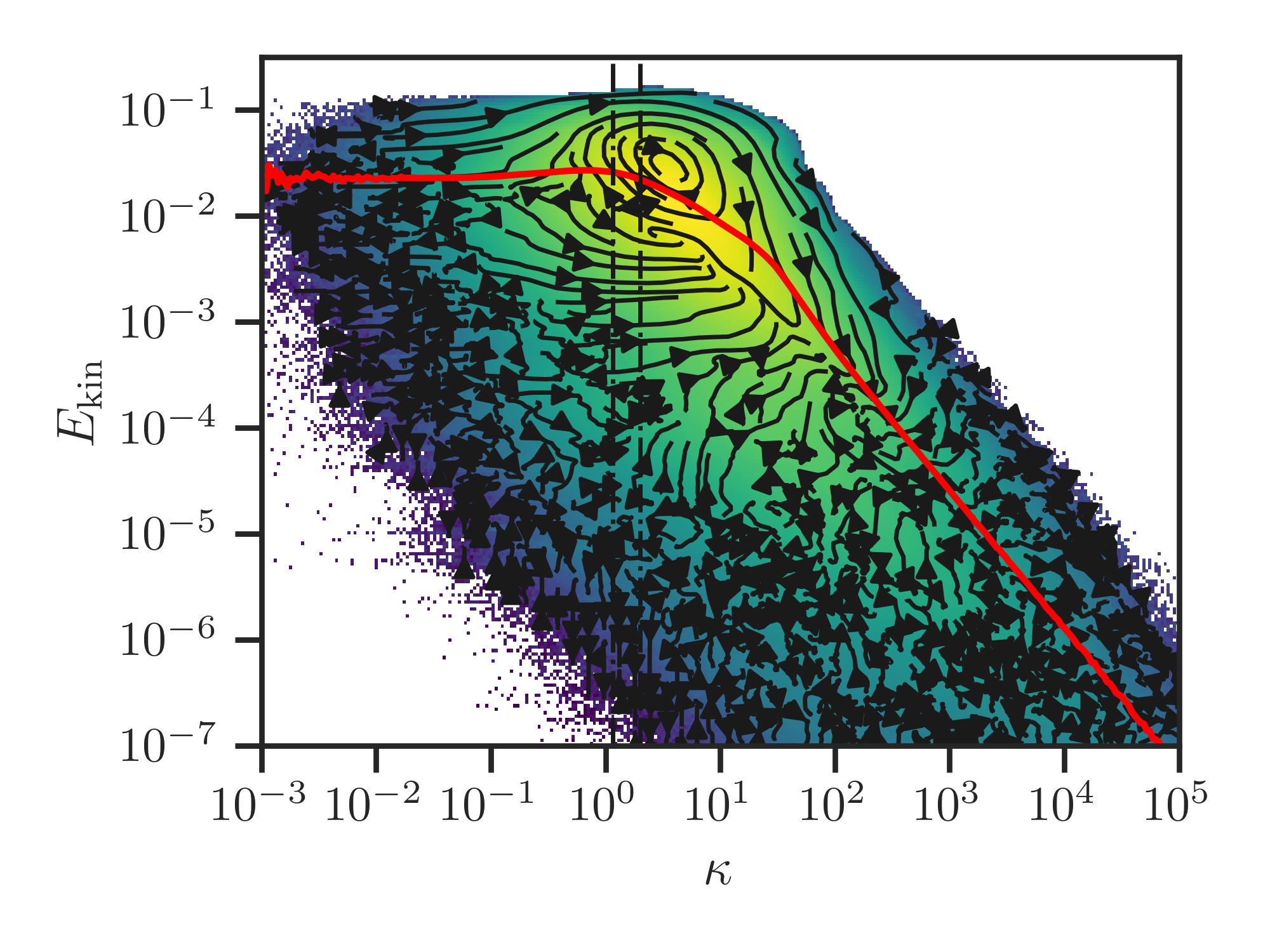} &
            \includegraphics[trim={0 0 0 0},clip, width=0.49\textwidth]{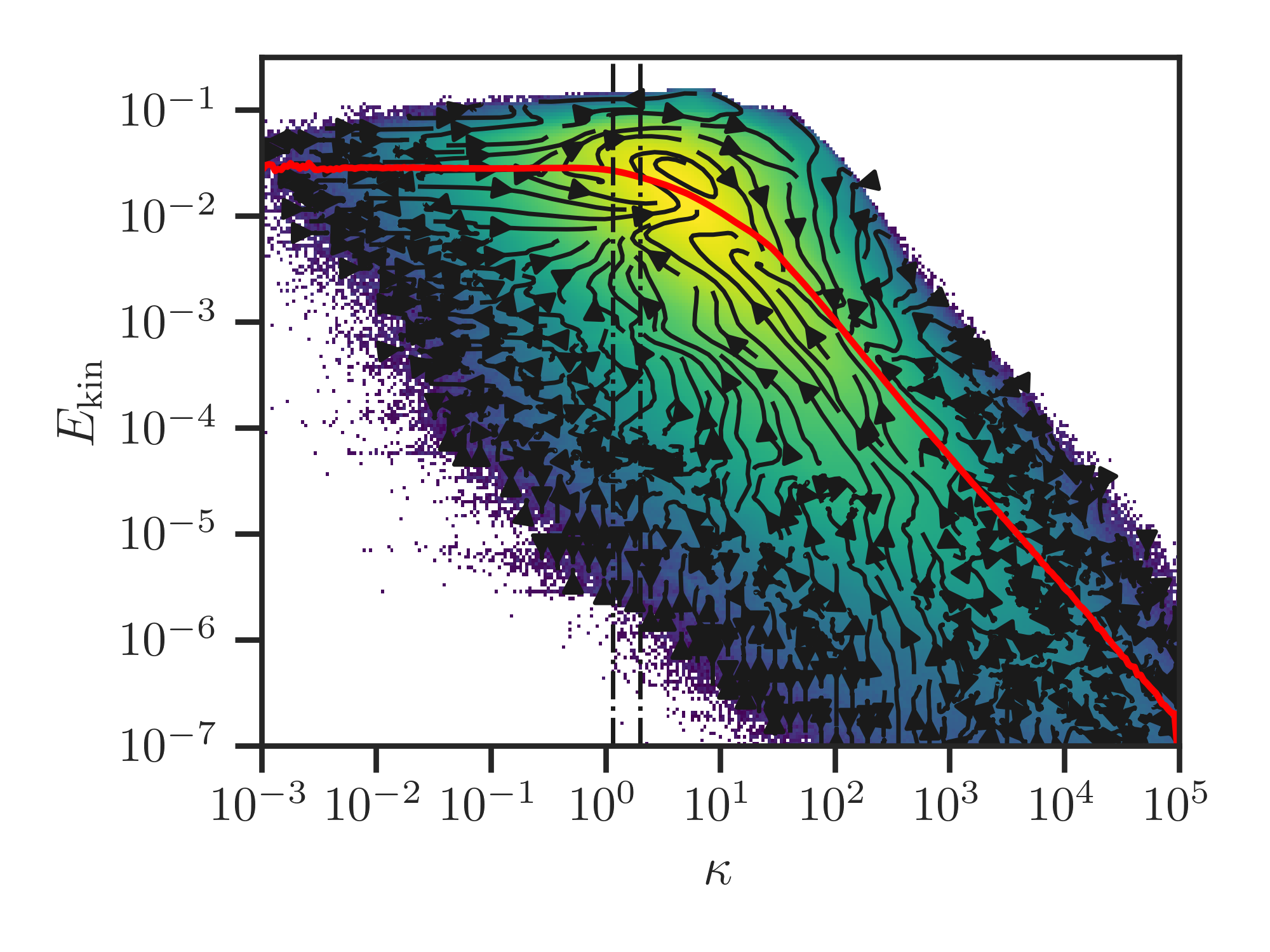}\\
        \end{tabular}
        
        \begin{tabular}{c}
             e) \\
            \includegraphics[trim={0 0 0 0},clip, width=0.49\textwidth]{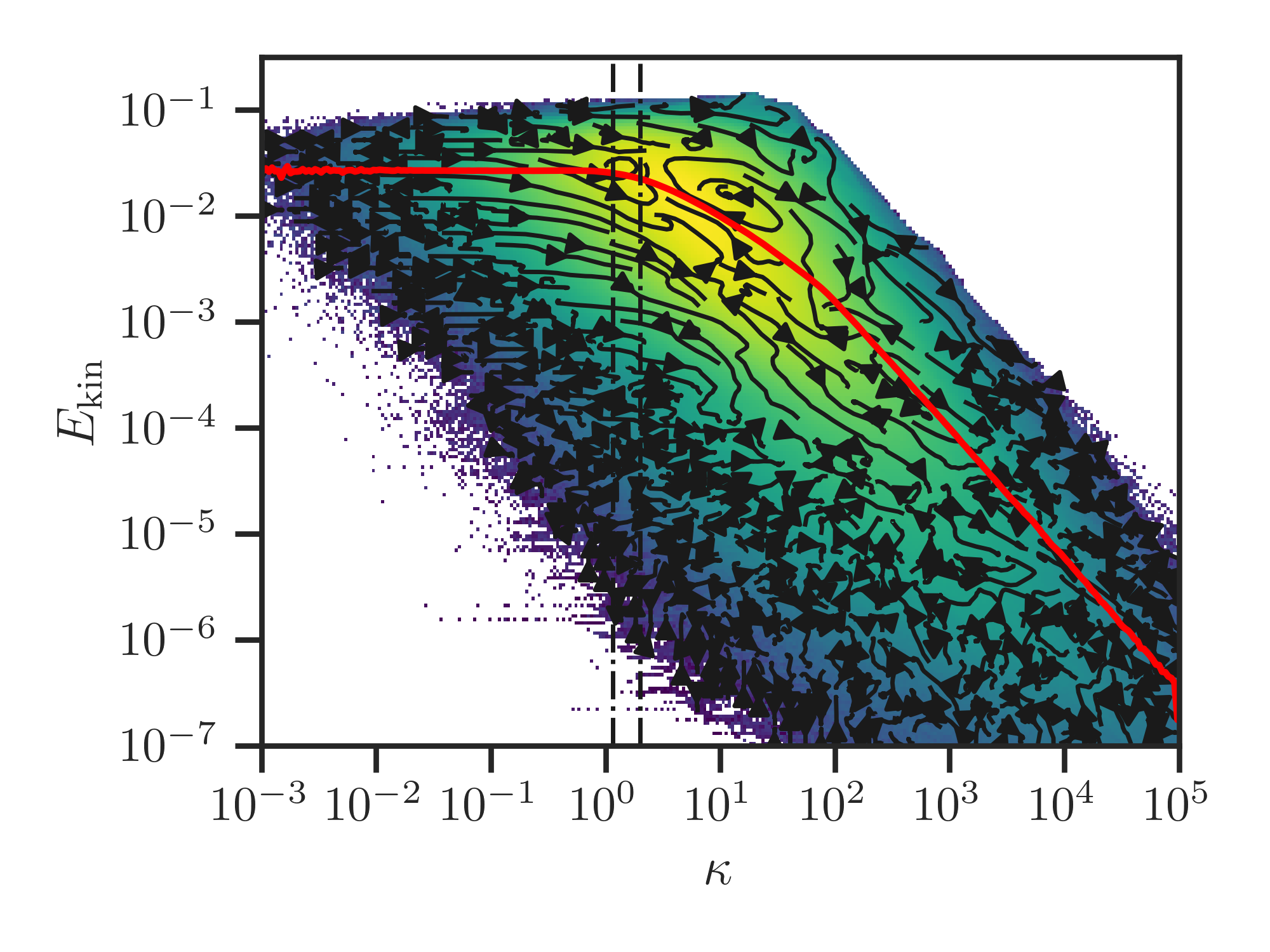}
        \end{tabular}
        \caption{Streamline visualisations for the time evolution vector $\boldsymbol{U}^*$ for various $\mathrm{Ra}$ numbers. The coloured backgrounds refer to the relative incidence, see Figure~\ref{fig:hit_timeevo}. a) $\mathrm{Ra}=5\times 10^5$ b) $\mathrm{Ra}=10^6$ c) $\mathrm{Ra}=10^7$ d) $\mathrm{Ra}=10^8$ e) $\mathrm{Ra}=10^9$.}\label{fig:Ra_streamlines}
    \end{center}
\end{figure}

Due to the pursued equal number of samples, the widely varying time intervals (see $N_t$ in Table~\ref{tab:DNS}) prohibit a comparison of the prevailing structures of $\boldsymbol{U}^*$ within the $E_\mathrm{kin}$-$\kappa$ plane.
Nevertheless, they are displayed in Figure~\ref{fig:Ra_streamlines} for the sake of completeness.
The existence of the displayed coherent structures implies a statistical dependence between the energy and curvature evolution of a fluid parcel and its current state.
In this respect, the circulations close to the maximum of the relative incidence for all cases except $\mathrm{Ra}=10^9$ are noteworthy.
However, their interpretation in terms of the Rayleigh number requires further studies with equal time intervals for each case.

\newpage

\bibliographystyle{elsarticle-num-names}
\bibliography{cascade}  




\end{document}